\documentclass[fleqn,usenatbib]{mnras}
\usepackage{newtxtext,newtxmath}
\usepackage{comment}
\usepackage[T1]{fontenc}
\usepackage{ae,aecompl}

\usepackage{graphicx}	
\usepackage{amsmath}	
\usepackage{amssymb}	
\usepackage{multirow}
\usepackage{color}
\usepackage[dvipsnames]{xcolor}

\usepackage{threeparttable, tablefootnote}
\usepackage{tabularx}
\usepackage{isotope}

\usepackage{etoolbox}
\makeatletter
\patchcmd\@combinedblfloats{\box\@outputbox}{\unvbox\@outputbox}{}{\errmessage{\noexpand patch failed}}
\makeatother
\def\reff@jnl#1{{\rm#1\/}}
\def\aj{\reff@jnl{AJ}}                 
\def\araa{\reff@jnl{ARA\&A}}           
\def\apj{\reff@jnl{ApJ}}               
\def\apjl{\reff@jnl{ApJ}}              
\def\apjs{\reff@jnl{ApJS}}             
\def\ao{\reff@jnl{Appl.Optics}}        
\def\apss{\reff@jnl{Ap\&SS}}           
\def\AAp{\reff@jnl{A\&A}}              
\def\AApr{\reff@jnl{A\&A~Rev.}}        
\def\AAps{\reff@jnl{A\&AS}}            
\def\azh{\reff@jnl{AZh}}               
\def\baas{\reff@jnl{BAAS}}             
\def\jcap{\reff@jnl{JCAP}}             
\def\jrasc{\reff@jnl{JRASC}}           
\def\memras{\reff@jnl{MmRAS}}          
\def\mnras{\reff@jnl{MNRAS}}           
\def\pra{\reff@jnl{Phys.Rev.A}}        
\def\prb{\reff@jnl{Phys.Rev.B}}        
\def\prc{\reff@jnl{Phys.Rev.C}}        
\def\prd{\reff@jnl{Phys.Rev.D}}        
\def\prl{\reff@jnl{Phys.Rev.Lett}}     
\def\pasp{\reff@jnl{PASP}}             
\def\pasj{\reff@jnl{PASJ}}             
\def\qjras{\reff@jnl{QJRAS}}           
\def\skytel{\reff@jnl{S\&T}}           
\def\solphys{\reff@jnl{Solar~Phys.}}   
\def\sovast{\reff@jnl{Soviet~Ast.}}    
\def\ssr{\reff@jnl{Space~Sci.Rev.}}    
\def\zap{\reff@jnl{ZAp}}               
\def\nat{\reff@jnl{Nature}}            

\def\snana{\textsc{snana}}

\defcitealias{2014AJ....147...99M}{M14}
\defcitealias{2014ApJS..213...19B}{B14}
\defcitealias{2017ApJS..233....6H}{H17}
\defcitealias{2014Ap&SS.354...89B}{Br14}
\defcitealias{2011MNRAS.412.1441L}{L11}
\defcitealias{2014AJ....147..118R}{R14}
\defcitealias{2017ApJ...843....6J}{J17}

\title[Spectrophotometric templates for core collapse SNe]
{Spectrophotometric templates for core collapse supernovae and their application in simulations of time-domain surveys}

\author[Vincenzi et al.]{M.~Vincenzi,$^{1,2}$\thanks{E-mail: maria.vincenzi@port.ac.uk} M.~Sullivan,$^3$
R.~E.~Firth,$^3$ C.~P.~Guti\'{e}rrez,$^3$ C.~Frohmaier,$^1$ M.~Smith,$^3$ \newauthor C.~Angus,$^3$  R.~C.~Nichol$^1$\\
$^1$Institute of Cosmology and Gravitation, University of Portsmouth, Portsmouth, PO1 3FX, UK\\
$^2$DISCnet Centre for Doctoral Training, University of Portsmouth, Portsmouth, PO1 3FX, UK\\
$^3$School of Physics and Astronomy, University of Southampton, Southampton, SO17 1BJ, UK}

\date{Version post referee report}

\pubyear{2019}

\begin{document}
\label{firstpage}
\pagerange{\pageref{firstpage}--\pageref{lastpage}}
\maketitle

\maketitle
\begin{abstract}
The design and analysis of time-domain sky surveys requires the ability to simulate accurately realistic populations of core collapse supernova (SN) events. We present a set of spectral time-series templates designed for this purpose, for both hydrogen-rich (type II, IIn, IIb) and stripped envelope (types Ib, Ic, Ic-BL) core collapse supernovae. We use photometric and spectroscopic data for 67 core collapse supernovae from the literature, and for each generate a time-series spectral template. The techniques used to build the templates are fully data-driven with no assumption of any parametric form or model for the light curves. The template-building code is open-source, and can be applied to any transient for which well-sampled multi-band photometry and multiple spectroscopic observations are available. We extend these spectral templates into the near-ultraviolet to $\lambda\simeq1600$\AA\ using observer-frame ultraviolet photometry. We also provide a set of templates corrected for host galaxy dust extinction, and provide a set of luminosity functions that can be used with our spectral templates in simulations. We give an example of how these templates can be used by integrating them within the popular SN simulation package \snana, and simulating core collapse supernovae in photometrically-selected cosmological type Ia supernova samples, prone to contamination from core collapse events.

\end{abstract}


\section{Introduction}\label{sec:intro}

Current and future time-domain optical surveys are expected to discover many thousands of optical transients and supernovae (SNe) on a nightly basis. 
This number far surpasses the resources available to spectroscopically confirm the nature of each detected event, and thus new approaches must be developed in order to exploit these samples. Techniques to model the composition of transients in future datasets are needed to guide and prioritise the limited follow-up resources, methods to classify transients based on their (possibly incomplete) photometric light curves are required to identify events of interest for different science goals, and accurate models of the expected contamination from core collapse SNe are needed for future cosmological samples of type Ia supernovae (SNe Ia).

Each of these tasks requires the ability to accurately model populations of core collapse SNe. For example, SN photometric classification techniques critically rely on templates of core collapse SNe, whether these are classical \lq template fitting\rq\ techniques \citep[e.g.,][]{2007ApJ...659..530K,2008AJ....135..348S,2011ApJ...738..162S,2009ApJ...707.1064R,2010ApJ...709.1420G},
or whether these are machine learning techniques \citep[][]{2012arXiv1201.6676I, 2013MNRAS.429.1278K, 2016ApJS..225...31L, 2019arXiv190106384M}
that must be trained on representative samples of data, often based on simulations.

A similar argument can be extended to SN Ia photometric cosmological analyses. The potential of cosmological analyses using photometrically-classified SNe Ia is significant and robust techniques will be required to unlock the power of future samples. In current SN Ia photometric cosmological analyses, the effect of contamination is assessed using simulation techniques \citep[][]{2007PhRvD..75j3508K, 2012ApJ...752...79H,2017ApJ...836...56K}, which are sensitive to how well the contamination can be modelled, rather than to the classification techniques used. The recent analysis of the photometric SN Ia sample from the Panoramic Survey Telescope and Rapid Response System \citep[Pan-STARRS 1;][]{2017ApJ...843....6J,2018ApJ...857...51J} has shown that simulations of core collapse SNe using currently available template libraries and luminosity functions significantly underestimate the apparent core collapse SN contamination in the data, and a significant tuning of the underlying properties of the core collapse SN population is required.

Historically, SN spectral templates have been constructed by combining together multiple observations of different SNe of the same type into some representative average spectral series of the class, interpolating between data to produce a spectral series sampled perhaps daily. This is particularly applicable to SNe Ia \citep{1991AAS...89..537L,2002PASP..114..803N,2007ApJ...663.1187H}, where the object-to-object diversity is relatively limited and quantifiable, and the concept of an average spectrum is straight forward. The advantage of this approach is that a spectrum can be interpolated on any epoch, from which photometry can be synthesised. However, this method is not as suitable for the far more heterogeneous core collapse SNe, where the spectral diversity is larger and the observed data typically sparser, thus making it more difficult (and perhaps conceptually meaningless) to combine data from different events into a single template.

This is well demonstrated by the complex classification scheme for core collapse SNe that depends on their spectral \citep[e.g.,][]{1997ARAA..35..309F} and sometimes photometric \citep[e.g.,][]{1979AA....72..287B,1994AA...282..731P,2012ApJ...756L..30A} properties; for a recent review see \citet{2016arXiv161109353G}. Type II SNe (SNe II) have clear signatures of hydrogen in their spectra whereas type I SNe do not; type Ib SNe (SNe Ib) have helium absorption lines whereas type Ic (SNe Ic) do not; and neither SNe Ib nor SNe Ic have the strong silicon or sulphur lines usually present in thermonuclear explosions. SNe Ic are often further divided according to whether they have broad lines (-BL) in their spectra, indicating very high velocities in the expanding ejecta. Type IIb SNe \citep[SNe IIb;][]{1993ApJ...415L.103F} are transition objects that begin as SNe II, but lose the hydrogen and develop increasingly strong helium lines as they evolve. Type IIn SNe \citep[SNe IIn;][]{1990MNRAS.244..269S} are SNe II that present narrow hydrogen emission lines, interpreted as the result of interactions between SN ejecta with circumstellar material \citep{1990SvAL...16..457C,2008ApJ...686..467S}. Thus, for core collapse SNe, a library of spectral templates, rather than a single average template, is more appropriate.

The first library of core collapse SN templates was developed for the Supernova Photometric Classification Challenge \citep[SNPhotCC;][]{2010arXiv1001.5210K,2010PASP..122.1415K}, with 41 templates based on  multi-band light curves of spectroscopically-confirmed core collapse SNe from the Carnegie Supernova Project (CSP), the Supernova Legacy Survey (SNLS), and the Sloan Digital Sky Survey-II (SDSS-II). In the absence of high-cadence spectral time series for these events, the same spectral energy distribution (SED) time series was used for each event, taken from the popular templates of Peter Nugent\footnote{\url{https://c3.lbl.gov/nugent/nugent_templates.html}}, and matched to the observed photometry of each individual event. Ultraviolet (UV) and near-infrared wavelengths for these templates are poorly constrained, and only observer-frame $U$/$u$-band photometry is included. This original library has been improved with additional events and is also used in the Photometric LSST Astronomical Time-Series Classification Challenge \citep[PLAsTiCC;][]{2018arXiv181000001T}.

Since the SNPhotCC challenge, the number of published core collapse SNe has significantly increased, with the release of large samples of spectroscopic \citep{2014AJ....147...99M, 2017ApJS..233....6H, 2018AA...609A.134S, 2019MNRAS.482.1545S} and photometric \citep{2012ApJ...756L..30A, 2013AA...555A..10T, 2014ApJS..213...19B, 2017ApJS..233....6H,  2017ApJ...850...89G,
2017ApJ...850...90G,
2018AA...609A.134S, 2018AA...609A.136T} data, alongside many single object studies of interesting or unusual events \citep[e.g.,][]{2011ApJ...728...14P,2017NatAs...1..713T,2017Natur.551..210A,2018Natur.554..497B,2018NatAs...2..574A,2019Natur.565..324I}. UV coverage of core collapse SNe has also improved, with observations from the \textit{Swift} satellite using the Ultra-Violet/Optical Telescope (UVOT) instrument \citep{2005SSRv..120...95R, 2009ApJ...700.1456B, 2014Ap&SS.354...89B, 2015JHEAp...7..111B}, and follow-up programs using the \textit{Hubble Space Telescope} and \textit{GALEX} \citep[e.g.,][]{2008ApJ...685L.117G, 2012ApJ...760L..33B, 2015ApJ...803...40B}. The construction of templates in the UV is essential for simulating transients at higher redshift, where the rest-frame UV is redshifted into the observer-frame optical.

Taking advantage of these new data, in this paper we present a method to construct a SN spectral template based on individual SN events for which well-constrained multi-band light curves and sparse spectroscopic observations have been measured. This technique is data-driven, and in principle can be generalised to any type of transient. Each final spectral template generated by our method consists of a daily sampled spectral time-series, extended into the UV, and optionally corrected for extinction due to dust in the SN host galaxy. Our goal is then to apply this general technique to all suitable core collapse SNe in the literature, providing a template library representing all suitable core collapse SNe. We also provide the general software implemented to build this library, allowing an easy expansion for SNe published in the future.

The paper is laid out as follows. A detailed description of our techniques is presented in Section~\ref{sec:pycoco}. 
In Section~\ref{sec:data}, we then select from the literature a sample of 67 well-observed core collapse SN events for use in constructing our templates, and discuss various challenges in using the data to build the templates. We then illustrate the use of our new library by combining with published estimates of SN rates and luminosity functions (Section~\ref{sec:lf_rates}), and simulating core collapse SN contamination in a large photometric SN survey (Section~\ref{sec:simulatons}). We present the results and the conclusions drawn from these tests in Section~\ref{sec:summary}. Where relevant, we assume a Hubble constant of $H_0=70$\,km\,s$^{-1}$\,Mpc$^{-1}$ and a flat, $\Lambda$CDM universe with a matter density of $\Omega_\mathrm{M}=0.3$.

\section{Method}
\label{sec:pycoco}

\begin{figure}
  \begin{center}
  \includegraphics[width = 0.475\textwidth]{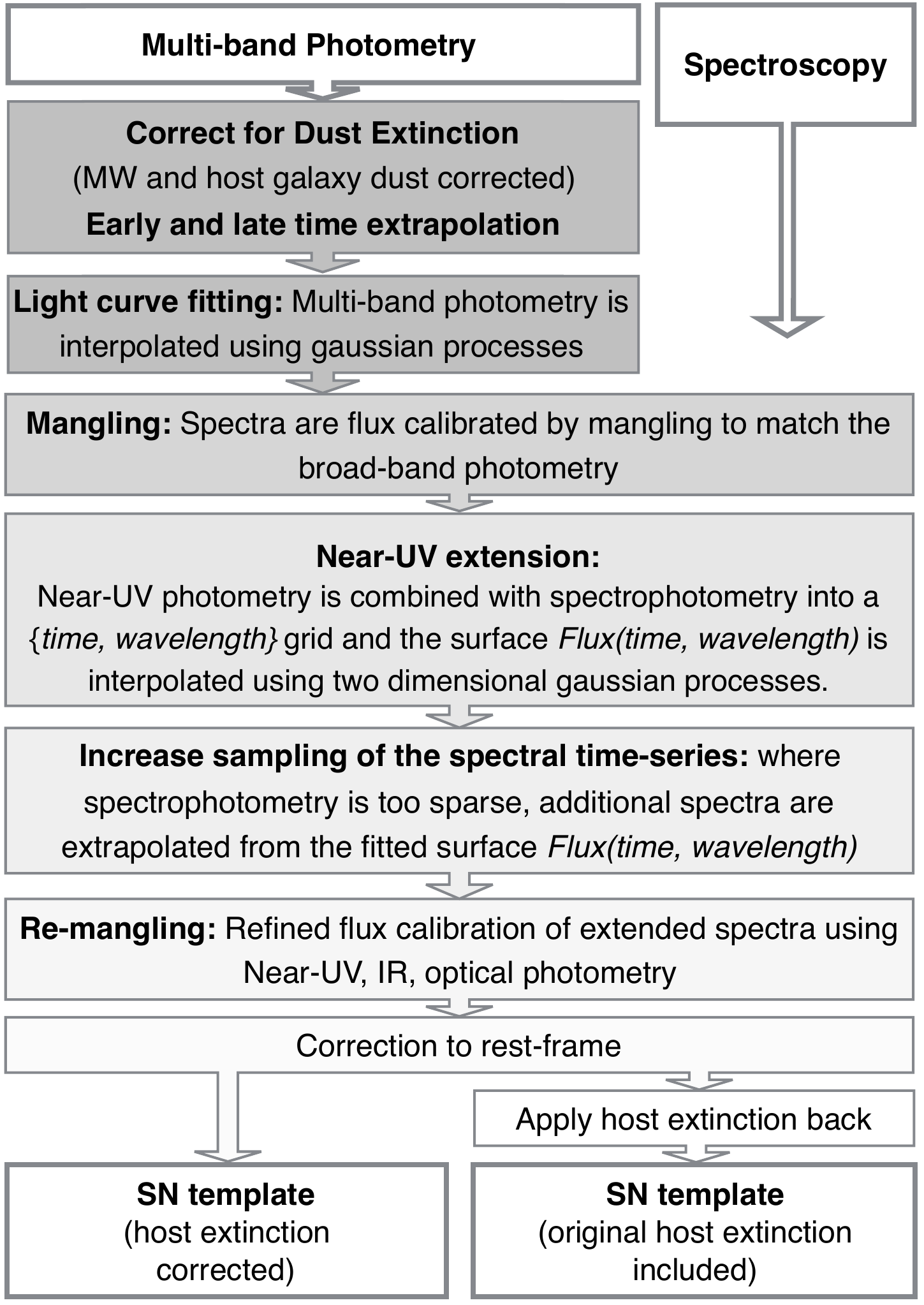}
  \caption{Schematic of the process used to build SN spectral templates from multi-band photometry and spectroscopy. The final templates are sampled daily and extended into the near-UV. The technique implemented to fit multi-band light curves is presented in Section~\ref{sec:light-curve-fits}, and the flux calibration and near-UV extension of the spectroscopic data are discussed in Section~\ref{sec:gener-spec-data}. In Section \ref{sec:final_tmpl} we explain the final corrections applied to produce the final SN template.}
  \label{FIG:PYCOCO}
  \end{center}
\end{figure}

In this section, we present the techniques designed to construct time-series spectral templates from photometric and spectroscopic observations of individual SN events. Fig.~\ref{FIG:PYCOCO} shows a schematic overview of the technique. (Our method assumes that any photometric data have been corrected for Milky Way and host galaxy extinction at the first order; we discuss this in detail in Appendix~\ref{Appendix:extinction} and when applying the code to data in Section~\ref{sec:data}.)

We first interpolate the observer-frame SN light-curves observed in multiple filters (Section~\ref{sec:light-curve-fits}), and we use this to estimate for each filter the photometric flux at the epochs on which spectral information for the SN is available. The interpolated photometry is then used to \lq flux-calibrate\rq\ the observed spectra by adjusting the overall spectral shape, but retaining the detailed spectral features (Section~\ref{sec:gener-spec-data}). We then use UV photometry from the same SN event to extend the spectral templates at bluer wavelengths (Section~\ref{sec:extension}). Finally, we smoothly interpolate between the (UV extended) calibrated spectra so that the final time series is sampled daily (Section~\ref{sec:final_tmpl}). We discuss each step in turn.

\subsection{Light curve fits}
\label{sec:light-curve-fits}

We explored various techniques to fit or model the SN observer-frame light curves, using different parametric forms. This included testing different models proposed in the literature \citep{2009AA...499..653B,2010PASP..122.1415K,2013MNRAS.429.1278K,2015ApJ...799..208S,1996ApJ...471L..37V}, and exploring different techniques to fit these parametric forms, including the \textsc{MultiNest} package \citep{feroz08, multinest} and the \textsc{mpfit} package \citep{2009ASPC..411..251M}. However, we found no single functional form that had both a sufficient flexibility to provide good fits across all filters for all core collapse SNe, but yet that could still be reliably constrained by typical SN data. For this reason, we opted instead to use a non-parametric data-driven interpolation technique, Gaussian processes. 

\subsubsection{Modelling light curves with Gaussian processes}
\label{sec:GPs-light-curves}

There are now several examples in the literature of the use of Gaussian processes \citep[GPs;][]{Rasmussen06gaussianprocesses} to interpolate transient light curves or spectral energy distributions \citep{2013ApJ...766...84K,2018ApJ...854..175I,2018arXiv181204071A, 2018ApJ...869..167S}. A GP can be seen as the generalisation of a multivariate Gaussian distribution. Formally, a GP generates data located throughout some domain (time, in the case of light curves) such that any finite subset of points in that domain follows a multivariate Gaussian distribution. As a multivariate Gaussian distribution is fully specified by its mean and covariance matrix, a GP is defined by a mean function and the covariance function, $k(x, x')$ (also called the kernel), that relates each point (of the light curve in this case) to any other. The hyperparameters needed to define the kernel are usually determined using a likelihood maximisation routine. 

Regression using GPs has several advantages. The first is that GP regression is a non-parametric model, and therefore has the flexibility to interpolate the wide heterogeneity of light-curve shapes inherent in core collapse SNe. Secondly, it is straight forward to include in the model the uncertainties of the observed photometric data points and, consequently, to estimate uncertainties on the interpolated light curve. However, while the \lq function-agnostic\rq\ nature of GP is a strength, the downside is that GPs are also physics-agnostic. As a consequence, negative fluxes, diverging and infinitely brightening light curves are, in principle, allowed. However, these non-physical behaviours typically occur when light curves are extrapolated outside the time-range covered by the photometric data, which can be avoided by restricting to interpolation only.

We use the python package \textsc{george} \citep{2015ITPAM..38..252A} to perform the GP regression on the light curves in flux space. We use a Matern 3/2 kernel function ($k$) of the form: 
\begin{equation}
    k(x, x') = A \left( 1 + \frac{\sqrt{3}(x-x')}{\sigma}\right) \exp{\left(-\frac{\sqrt{3}(x-x')}{\sigma}\right)},
    \label{eq:matern}
\end{equation} 
where $A$ is an amplitude factor, $\sigma$ regulates the scale at which correlations are significant, and $x$ is the domain over which the regression is being performed, which in this case is time. Generally we optimise the free hyperparameters of the model (amplitude and scale of the kernel function) by minimising the log-likelihood of the model given the data, and we set the prior on the mean function to a constant zero function.

\subsubsection{Early and late phases}

The GP interpolation works well at most light-curve phases, but requires extra attention at the start and end of the light curve (where photometric sampling is sparser or absent) to ensure that the interpolation is robust.

At late phases, the SN luminosity is powered by $^{56}$Co decay. Sparse photometric measurements are sufficient to constrain the SN evolution in this phase, and so SNe are typically observed less frequently. However, when the data become sparser, the GP interpolation  becomes less informative, and with larger uncertainties. We therefore visually inspect each light curve and determine at which phase it becomes dominated by $^{56}$Co decay. We then interpolate/extrapolate additional photometric points and propagate their uncertainties from the linear fit for use in the GP interpolation. This \lq oversampling\rq\ of the light curve significantly improves the GP interpolation, without adjusting the original data around peak.

At early phases we use a similar approach and over-sample the rising part of the light curve, filling the gap between the estimated explosion day and the first photometric data point. We fit the early data flux $f$ as a function of time $t$ with a widely used parametrization of
\begin{equation}
  f(t)=\begin{cases}
    \alpha(t-t_0)^n & \text{if $t\geq t_0$}\\
    0 & \text{if $t<t_0$},
  \end{cases}
\end{equation}
where $\alpha$ is a normalisation coefficient, $t_0$ is the time of explosion, and $n$ is the power index of the rising light curve.

For most of the SNe, an accurate estimate of $t_0$ can be found in the literature, typically derived from light-curve modelling or from the analysis of early spectroscopic observations \citep[e.g.,][]{2016ApJ...821...57D}. Thus, we generally fix the parameter $t_0$. When the explosion epoch is uncertain, we treat $t_0$ as a free global parameter and use non-detections and the SN date of discovery as lower and upper bounds on $t_0$. The power index $n$ is fixed only when the available early photometry is too poor to constrain the fit. For stripped envelope SNe, we assume a power law of the form $f(t)\propto t^{1.5}$, expected for the cooling of the shock-heated expanding ejecta \citep[][]{2013ApJ...769...67P}. For hydrogen-rich SNe, we assume$f(t)\propto t^{0.935}$, \citep{2015MNRAS.451.2212G}.

For some SNe considered (SN\,1987A, SN\,1993J, SN\,2006aj, SN\,2013df, SN\,2011dh, SN\,2011fu) an initial shock-breakout is detected in the light curves, presenting as an initial peak in the photometry. In such cases, the fitted functional form used is:
\begin{equation}
  f(t)=\begin{cases}
    \alpha(t-t_0)^n + \mathcal{N}(t_{b}, \sigma_{b}) (1-e^{t-t_0}), & \text{if $t\geq t_0$}.\\
    0, & \text{if $t<t_0$}.
  \end{cases}
\end{equation}
where $t_{b}$ and $\sigma_{b}$ are the peak and width of the initial bump in the light curve.

\begin{figure*}
\includegraphics[width=0.97\textwidth]{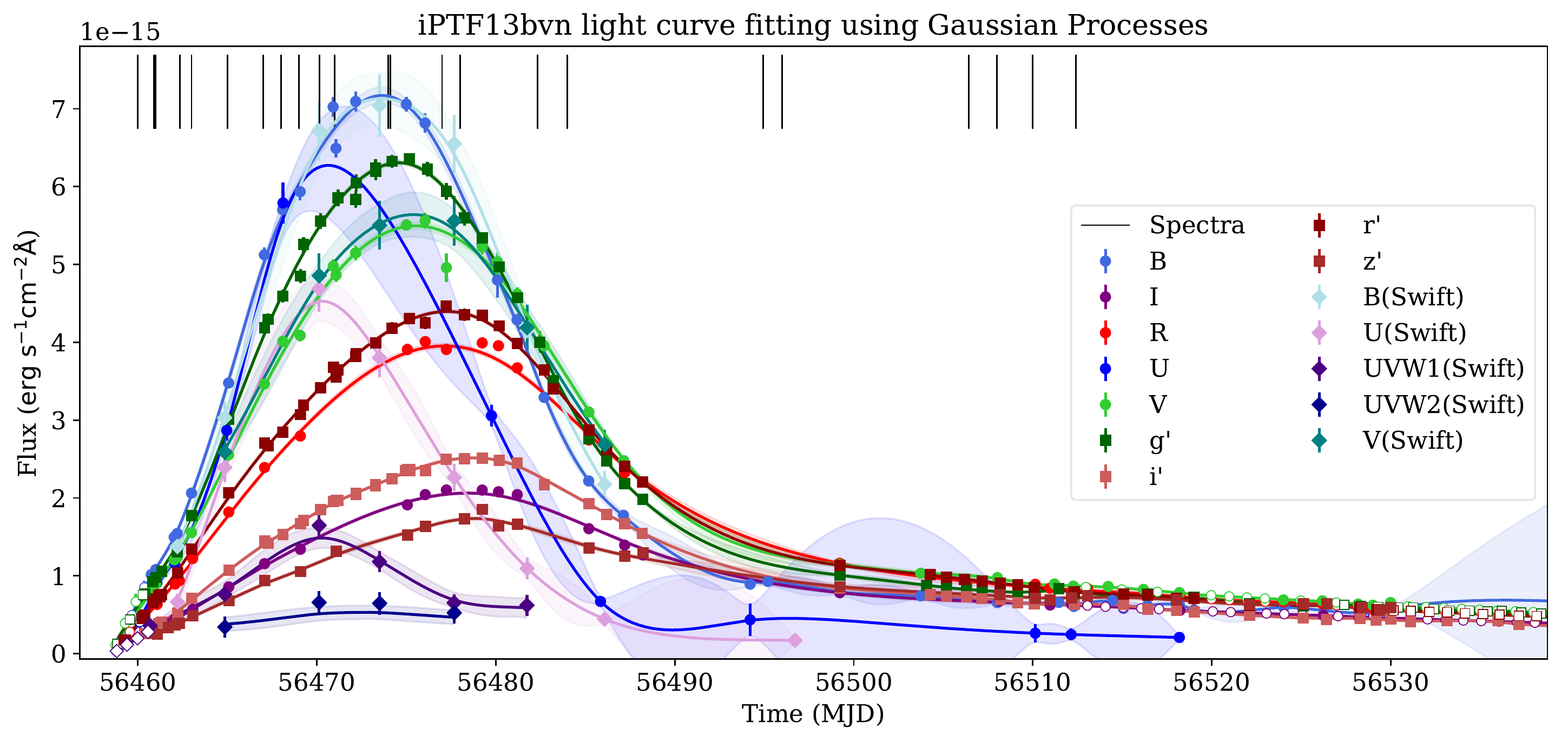}\\
\caption{Fit of multi-band light curves using Gaussian processes for the SN Ib iPTF13bvn. iPTF13bvn was observed in 14 filters, including five filters from \textit{Swift}/UVOT. Filled dots indicate the published photometry. The cobalt decay dominated phase of each light curve is artificially over-sampled to better constrain the GP interpolation at later phases, denoted with open symbols (see Section~\ref{sec:data_prep}). Solid lines show the interpolation estimated using Gaussian processes, and the shaded area shows the uncertainty on this interpolation. The black vertical lines indicate epochs at which spectra were obtained.}
\label{FIG:PTF_LCfit}
\end{figure*}

Fig.~\ref{FIG:PTF_LCfit} shows an example of a multi-colour light curve interpolation using GPs for the SN iPTF13bvn \citep[][]{2014AA...565A.114F}. Equivalent figures for all other events in our sample are available as online-only figures.

\begin{figure*}
\includegraphics[width=0.97\textwidth]{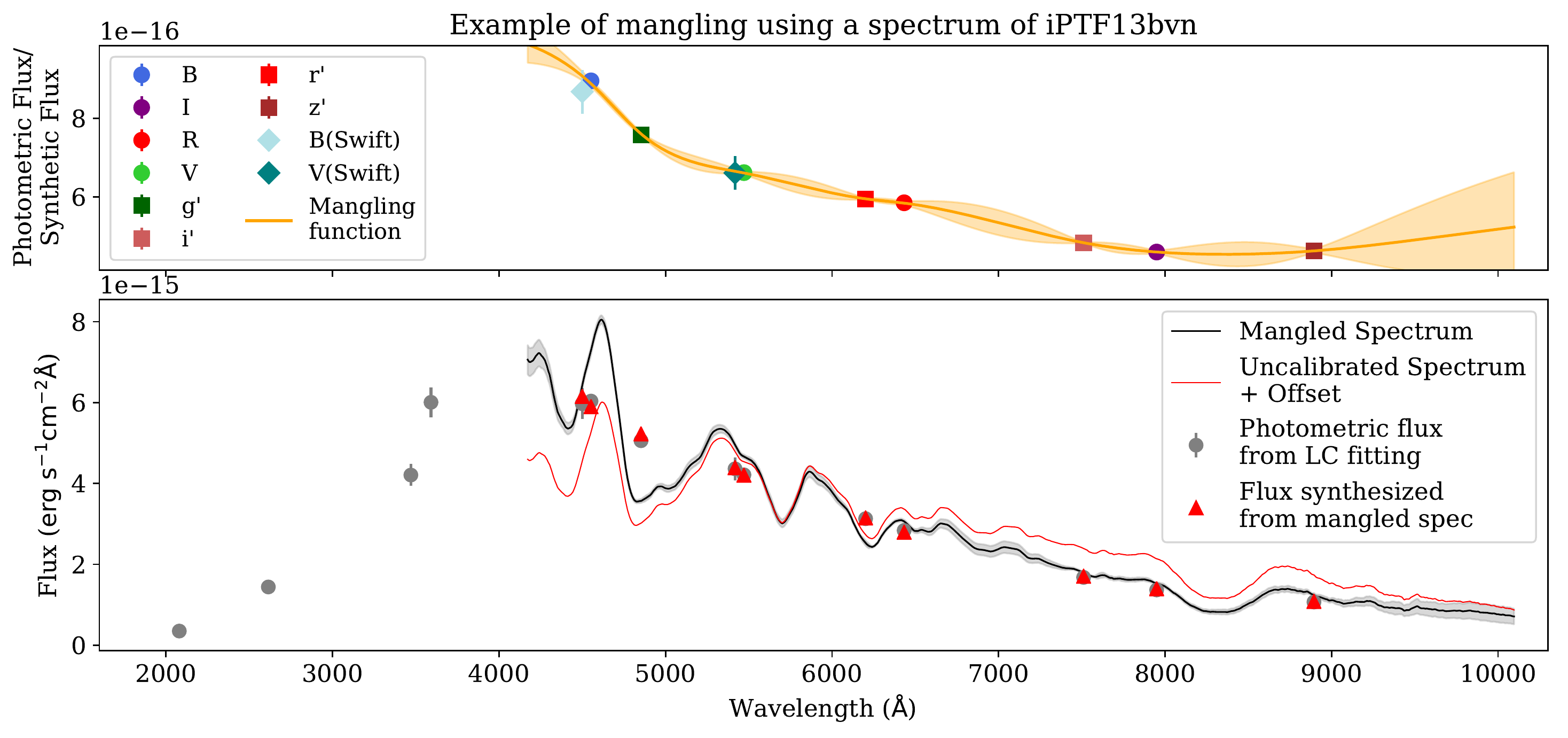}\\ 
\caption{Mangling of the iPTF13bvn spectrum taken at MJD 56469.0. \emph{Top panel}: For each filter we show the ratio between the photometric flux (estimated using the GP interpolation of the light curve at the epoch of spectral observation) and the synthetic flux from the observed spectrum. The uncertainties from the light curve GP interpolation are propagated. The smooth mangling correction function is also interpolated using GPs (solid orange line). \emph{Lower panel}: The comparison between the observed spectrum (red line) and the \lq flux-calibrated\rq\ spectrum following the mangling technique (black line). The photometric fluxes estimated from the light-curve interpolation are shown for each filter as grey dots. The fluxes synthesised from the flux-calibrated spectrum are also shown (red symbols) for comparison. Given the limited wavelength coverage of the spectrum in the near-UV, the photometry in the bluer filters is not used in the mangling process, but is used later to extend the spectrum into the near-UV.}
\label{FIG:PTF_mangling}
\end{figure*} 

\subsection{Generating spectrophotometric data}
\label{sec:gener-spec-data}
Observed SN spectra are rarely spectrophotometric, no matter how carefully the observations and reductions are performed. Differential slit losses during the observation, centring errors of the object on the slit, or non-photometric observing conditions not only affect the overall normalisation of the spectrum, but can also introduce a smooth, wavelength-dependent distortion of the continuum \citep{2013AA...549A...8B}.

Multi-band photometry allows a correction to be made for many of these effects: it can set the overall flux normalisation, and also provides information on the broad-band colours of the spectrum. We therefore adjust the observed spectra so that they match, in colour space, the photometry interpolated at the epoch of spectral observation. This correction is sometimes referred to as \lq mangling\rq\ the spectrum \citep[e.g.,][]{2007ApJ...663.1187H,2008ApJ...681..482C}, and consists of determining a smooth wavelength-dependent function that, when multiplied by the original spectrum, produces a spectrum with the correct colours and normalisation. In order to calculate this smooth wavelength-dependent function we again use GPs. 

For each SN spectrum, we first calculate the observed flux in each of the filters for which a photometric light curve is available. We compare these synthetic fluxes with the photometry interpolated from the light curve (Section~\ref{sec:light-curve-fits}), and determine the ratio of the two, propagating the uncertainties from the GP light curve interpolation and from the spectrum. We place each calculated ratio at the filter effective wavelength of the original spectrum, and use GPs to interpolate as a function of wavelength, obtaining a smooth wavelength-dependent calibration function. We use again a Matern 3/2 kernel (Eq.~\ref{eq:matern}), fixing the scale to 300\,\AA.

We do not use the common approach of using spline interpolation in the mangling. We explored this approach in detail, but found it is difficult to account correctly for the observational uncertainties in the photometry. The spline interpolation can also produce discontinuities in the mangling function for SNe for which photometry in neighbouring filters with close effective wavelengths is available (e.g., SNe which have photometry in both SDSS $r$ and Bessell $R$). Fig.~\ref{FIG:PTF_mangling} shows an example of a mangled spectrum and the interpolated mangling function.

\subsubsection{Extending the wavelength coverage of the spectra}
\label{sec:extension}

During the mangling procedure described above, any near-UV or near-infrared (near-IR) data from the SN are not used, as the spectroscopic data typically cover the optical wavelength range\footnote{Less then 10 per cent of the spectra considered in Section~\ref{sec:data} have coverage below 4000\,\AA\ and only 50 per cent cover $>$9000\,\AA.}. However, near-UV coverage in the SN rest frame is critical to simulate SN events at higher redshift (i.e., typically at $z>0.2$). We therefore use near-UV photometry observed for the same SN to smoothly extrapolate the spectra to lower wavelengths.

We experimented with smoothly extrapolating the spectrophotometric data by fitting a black body function to the optical and near-UV photometry. While this works for the rather featureless very early spectra of some SN subtypes, it provides a poor near-UV representation for most of the SN events we consider. The complex absorption due to ionised metals present in the SN envelope typically dominates the UV part of the SN SED during most of its evolution, and its modelling is non trivial \citep{2008ApJ...685L.117G, 2009ApJ...700.1456B, 2012ApJ...760L..33B, 2015ApJ...803...40B}. We therefore decided to use a less rigid approach and again implement GPs. In this case, we combine photometry (i.e., integrated flux versus time, $t$) and spectrophotometry (i.e., flux density versus wavelength, $\lambda$) into a two-dimensional grid on which we can interpolate a flux surface $f(t,\lambda)$. 

The implementation of a two-dimensional interpolation of this flux surface $f(t,\lambda)$ presents several advantages. First, it guarantees that the near-UV extension is continuous in time; extending each spectrum individually can result in large discontinuities in the final spectral time series in the near-UV.  In addition, it is common to have spectroscopic observations of the same SN from different facilities. As a result, the wavelength coverage of the available spectra can change significantly from spectrum to spectrum, with some spectra extending to $\sim3200$\,\AA, while other ending at 4000-4500\AA. A two-dimensional interpolation allows us to perform an extension including any information from adjacent spectra, and provides a final template that is continuous in time. Moreover, a two-dimensional model not only allows us to extend the spectrophotometry into the near-UV or near-IR, but also to simultaneously smoothly interpolate between the spectra and extrapolate \textit{additional} spectra when the available data are too sparse (see Section~\ref{sec:final_tmpl}).

\begin{figure*}
 \centering
 \includegraphics[width=0.9\textwidth]{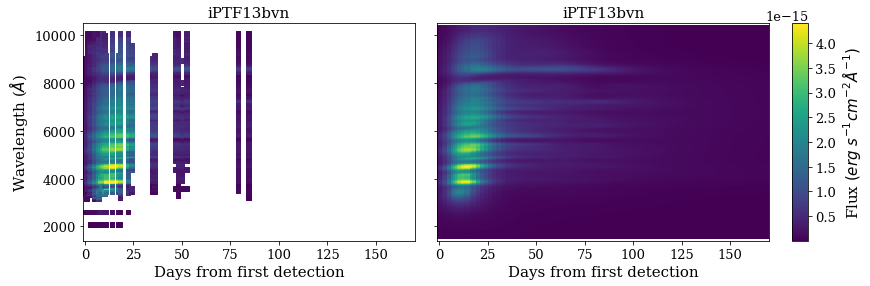}\\
 \caption{\emph{Left}: Two dimensional surface built for the SN Ib iPTF13bvn, combining sparse flux-calibrated spectroscopy and near-UV photometry as described in Section~\ref{sec:extension}. For this SN, photometry in the \textit{Swift} filters $u$, $uvw1$ and $uvw2$ is available. In this graphical representation of the two-dimensional flux surface $f(t, \lambda)$, spectrophotometry is represented by continuous vertical stripes, and light curve photometry by horizontal stripes. \emph{Right}: The smoothed interpolation of the flux surface $f(t, \lambda)$ surface using two-dimensional Gaussian processes.}
 \label{FIG:example_grid}
 \end{figure*}

In order to interpolate the flux surface $f(t, \lambda)$, both photometric and spectrophotometric data are combined and smoothed on a grid with a 60\,\AA\ binning in wavelength (to uniformly sample all the calibrated spectra and reduce the number of data points used to train the GP model). An example is shown in Fig.~\ref{FIG:example_grid}. The UV integrated flux is included and placed on the wavelength axis at the mean wavelength of the filter. The kernel used for GP interpolation is a two-dimensional Matern 3/2 Kernel function (Eq.~\ref{eq:matern}), where $x$ is now the two-dimensional vector $\mathbf{x} = \{x_{t}, x_{\lambda}\}$. The GP hyper-parameters, $\sigma_{\lambda}$ and $\sigma_{t}$ are fixed to 100\AA\ and 30 days respectively (optimising these hyper-parameters does not affect the results significantly and is computationally expensive). 

The two-dimensional GP interpolation requires a prior for the mean function to prevent non-physical behaviour (i.e., negative fluxes or a GP model rapidly decreasing to zero when the data coverage is poor) from occurring. Building this prior is not trivial given the diversity of core collapse SNe in terms of brightness, shape and colours. In order to do this we use observed data from the 67 core collapse SNe that we will introduce in Section~\ref{sec:data}, examining their colour evolution in the optical and near-UV after dust extinction corrections (these SNe are all at $z<0.03$ where $k$-corrections are small). To first order, SNe II and fast rising SNe IIn (hydrogen-rich SNe) present a similar colour evolution, despite having different absolute brightnesses. The same is true for the stripped envelope SN classes of SNe IIb, SNe Ib and SNe Ic/Ic-BL \citep[see also][]{2011ApJ...741...97D, 2018AA...609A.135S} and for the sub-class of slow rising SNe IIn like SN\,2006aa and SN\,2011ht.

First, we measure, for each SN and for each optical and near-UV filter $X$, the evolution of $F_X(t)/F_V(t)$, where $F_X$ and $F_V$ are the GP interpolated light-curves in filters $X$ and the $V$-band respectively. We then average $F_X(t)/F_V(t)$ over all SNe within the same sub-class (hydrogen rich, stripped envelope or slow SNe IIn). We also measure, in each filter, the average wavelength weighted by a typical hydrogen-rich, stripped-envelope or slow Type IIn SED. We interpret each average colour evolution $\langle F_X(t)/F_V(t) \rangle$ as a monochromatic measurement of colour at the respective average wavelength, and we smoothly reconstruct the \lq colour surface\rq~$\langle F_{X-V}(t,\lambda)\rangle$. For each SN, the prior used to interpolate the flux surface $f(t, \lambda)$ is calculated multiplying the \lq colour surface\rq~$\langle F_{X-V}(t,\lambda)\rangle$ by the $V$-band light curve of the SN itself. This means that for each SN we build a different prior that has the colour properties of the sub class of SNe it belongs to, but is still normalised to the apparent brightness of the SN considered.

We test the effects of our prior using some of the best observed SNe (e.g., SN\,2013by and iPTF13bvn). We perform the two-dimensional GP interpolation with and without the prior, using the full set of data available for the SN, and then removing some information (i.e., simulating reduced UV coverage or sparser spectroscopy). We find that the choice of the prior improves the results when sparser data or less UV coverage is available, without significantly overestimating or underestimating the final near-UV extension, and without affecting the accuracy of the spectrophotometry.

\begin{figure*}
\begin{tabular}{cc}
\includegraphics[width=0.975\textwidth]{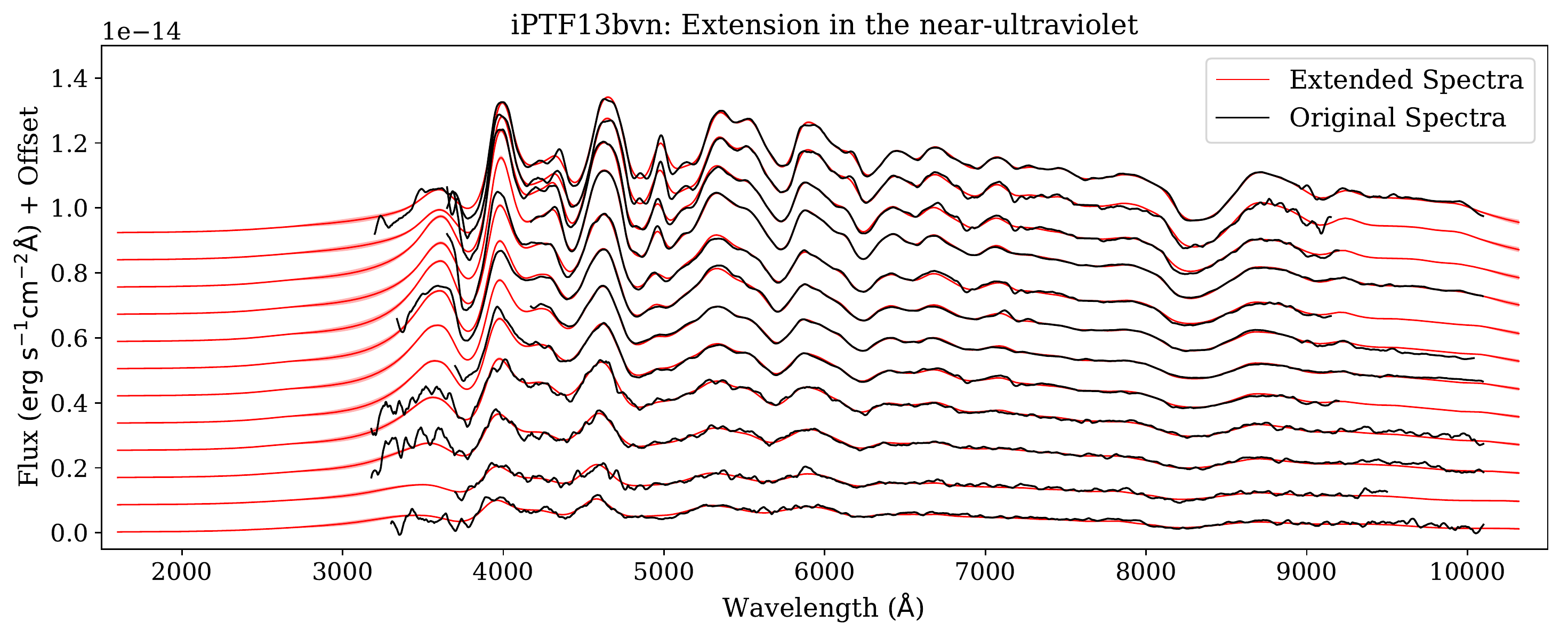}\\
\end{tabular}
\caption{Near-UV extension of spectrophotometric observations for the SN Ib iPTF13bvn. Each flux-calibrated spectrum (black) is extended using near-UV light curves and the two-dimensional interpolation technique presented in Section~\ref{sec:extension}. The near-UV extended spectra are shown in red.}
\label{FIG:PTF_extension}
\end{figure*}

Once the prior is built, the flux surface $f(t, \lambda)$ is interpolated and it is then straight forward to extrapolate for each optical spectrum the near-UV extension. 
 Fig.~\ref{FIG:PTF_extension} shows an example of spectra observed for the SN Ib iPTF13bvn and extended applying the technique presented in this Section. In Fig.~\ref{FIG:example_grid} we show, for the same SN, a graphical representation of the flux surface $f(t, \lambda)$ as reconstructed from the two-dimensional GP.

\subsubsection{Refining the calibration of the extended spectra}
\label{sec:remangle}

The extended spectra reconstructed from the two-dimensional interpolation provide a good approximation of the underlying near-UV SED. However, when building the two-dimensional surface, we interpret the photometric measurements as monochromatic flux density measurements at the filter mean wavelength. Ideally, the effective wavelength -- $\lambda_\mathrm{eff}$, the average wavelength of the filter weighted by the SED -- should be used instead, but this is not known a priori as the SED is not known. This can lead to inaccuracies, particularly in the near-UV when using photometry from \textit{Swift} \citep{2016AJ....152..102B}, where the filter transmission functions present significant tails into the optical. This can be relevant for some core collapse SNe with strong UV variability or at very early phases. As a consequence, the (time dependent) filter effective wavelength can be significantly different from the filter mean wavelength as the SED changes, and as a result, fluxes synthesised from our extended spectra in the \textit{Swift} filters can be up to 40-50 per cent discrepant when compared to the observed photometric fluxes.

We address this by adjusting the near-UV fluxes to remove the optical contribution from the filter red leaks before including them in the two-dimensional grid for interpolation. We do this by integrating the spectra over each filter's optical tail, subtracting from the original \textit{Swift} photometry, and then using this subtracted photometry in the two-dimensional grid for the interpolation. After the spectral time series is extended, we then apply a further correction to the final extended spectra. For each extended spectrum, we compute synthetic photometry in the near-UV and optical filters, and compare with the original photometry observed in the same filters. We then adjust the extended spectra until the synthetic and observed photometry match. We apply this process iteratively as $\lambda_\mathrm{eff}$ for each filter changes as the spectrum is mangled. We refer to this additional correction as \lq remangling\rq. At the end of this process, the extended spectra are robustly flux-calibrated at all wavelengths.

During the remangling, the GP model provides not only a prediction of the mangling function, but also the relative full covariance matrix. This includes the uncertainties on the input photometry (added in quadrature to the diagonal of the covariance matrix of the GP model) and provides a robust estimate of the (highly correlated) uncertainties on the final flux calibrated spectrum.

\subsection{Final templates}
\label{sec:final_tmpl}

The most widely used SN simulation packages usually simulate SN light curves by synthesising broadband photometry from a spectral time series model, and then linearly interpolating between this synthetic photometry to generate a light curve at any required cadence. For this reason, the optimal time sampling for our final time series SED templates is approximately one or two days, for a total phase coverage from a few days prior to maximum brightness to around a hundred days after maximum brightness. The observed photometry for our SN events typically meet this quality of sampling. However, the spectroscopic observations generally have a significantly lower cadence and poorer phase coverage.

We therefore increase the time sampling of our templates by smoothly interpolating between the calibrated spectra. This smooth interpolation is performed as described in Section~\ref{sec:extension}, and an example is shown in Fig.~\ref{FIG:example_grid}. Additional spectra are extrapolated from the interpolated flux surface and then remangled as described in Section~\ref{sec:extension}, ensuring that the colours and normalisation of the synthetic extrapolated spectra data are accurate.

As a final step, we transform each template to the rest-frame using the measured heliocentric redshift. When necessary, corrections for peculiar velocities are applied. Each SN template is daily sampled and its phase is calibrated arbitrarily by defining day zero as the day at which the synthesised $BVRI$ pseudo-bolometric light curve reaches its maximum value.

\begin{figure*}
\centering
\includegraphics[width=0.475\textwidth]{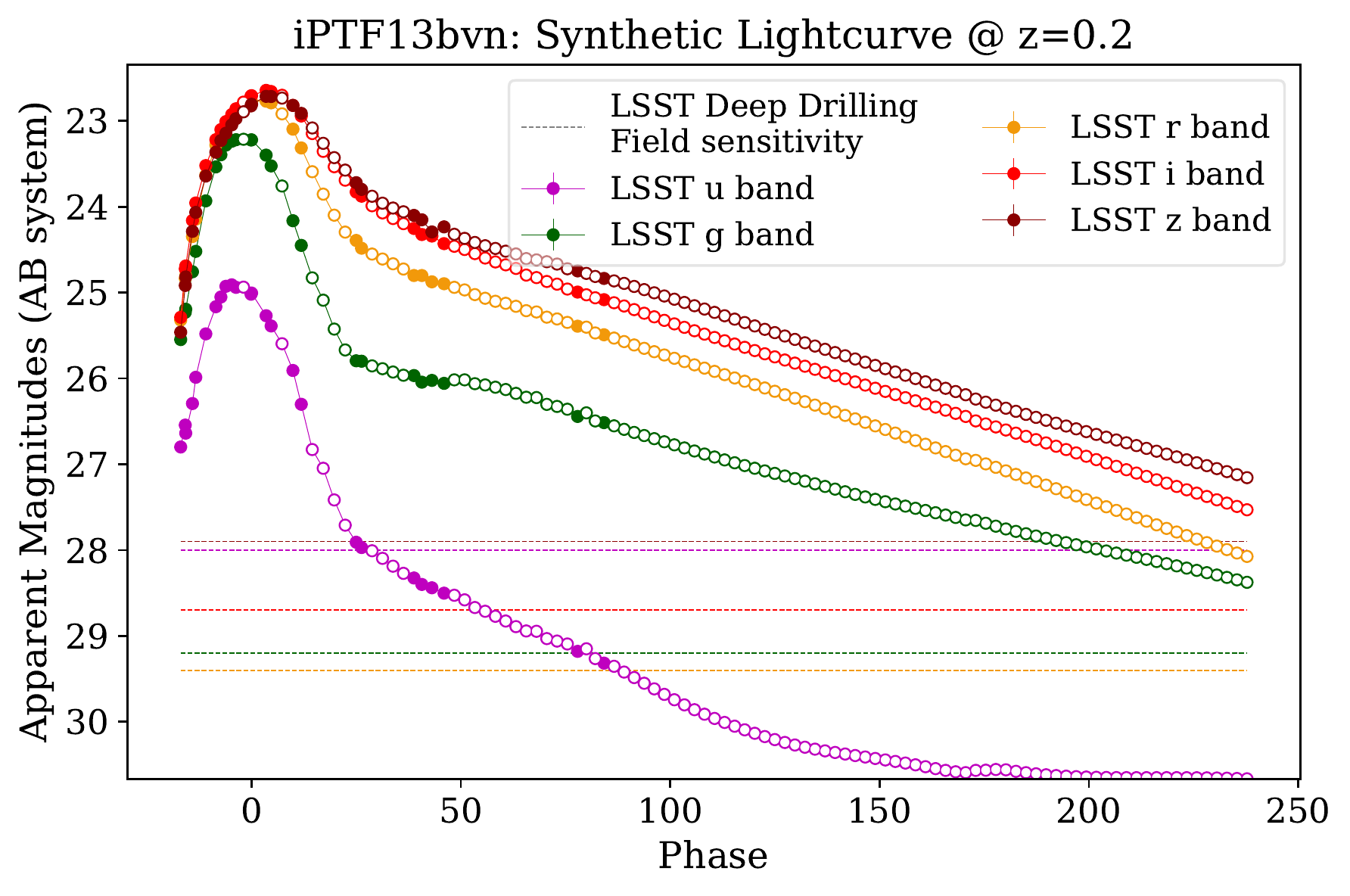}
\includegraphics[width=0.475\textwidth]{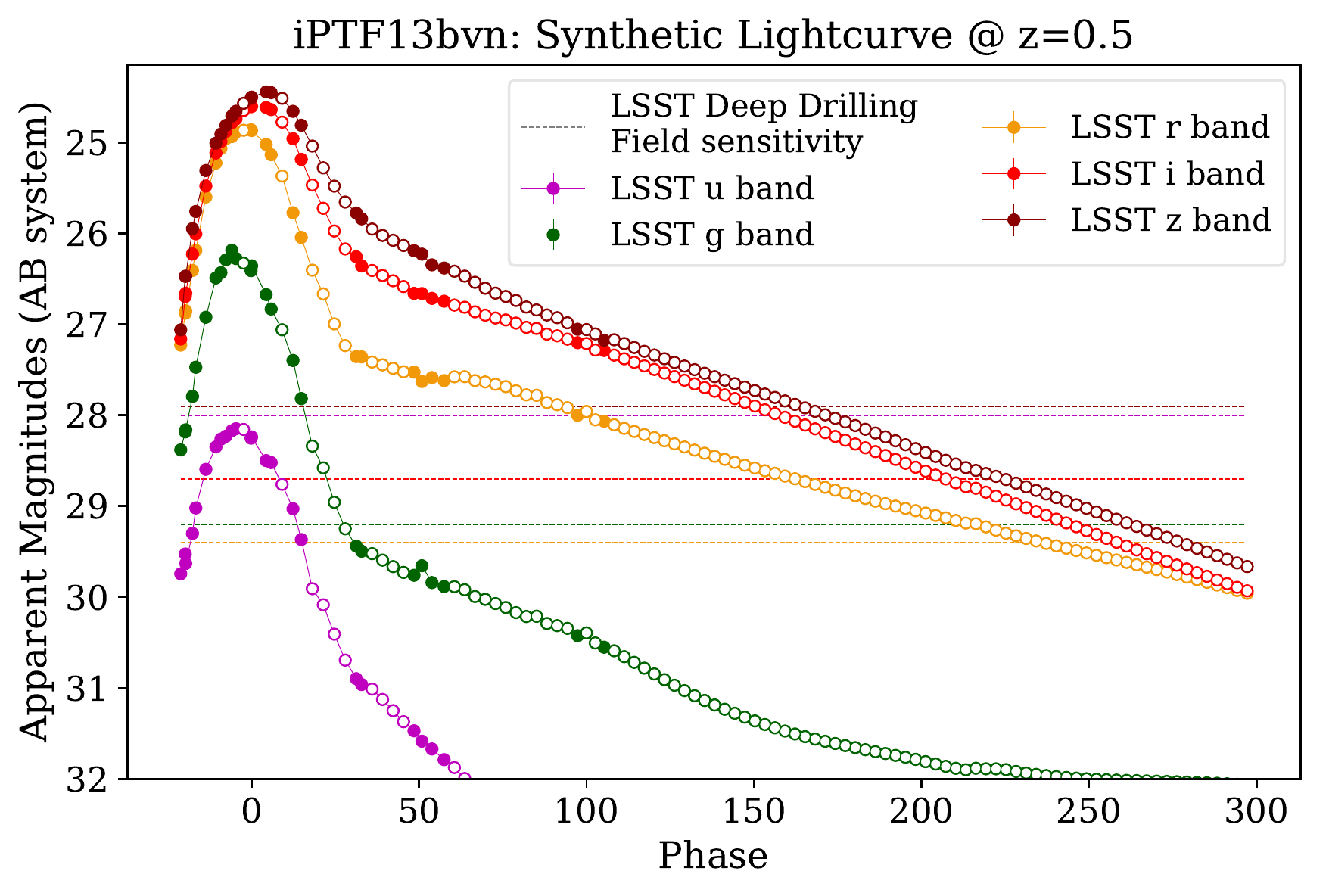}
\caption{Synthetic multi-band photometry estimated using the SN template built from the SN Ib iPTF13bvn. The photometry is synthesised in the LSST filter system, with the SN event simulated at $z=0.2$ (left panel) and $z=0.5$ (right panel). Filled symbols indicate photometric points synthesised from calibrated and extended spectroscopic observations; open symbols indicate photometric points synthesised from spectra extrapolated from the flux surface $f(t,\lambda)$ (see Section~\ref{sec:final_tmpl}). For comparison, we show for each filter the sensitivity estimated for the LSST Deep Drilling Fields. Equivalent figures for all other events in our sample are available as online-only figures.}
\label{FIG:LC_example_iPTF}
\end{figure*}

The result is a daily sampled and de-reddened, rest-frame spectrophotometric template. These templates can be used for simulations of SN light curves up to $z\approx 1$, in any optical filter system and assuming any Milky Way or host galaxy dust reddening.  We also provide templates for which host-galaxy extinction has not been corrected. In Fig.~\ref{FIG:LC_example_iPTF}, we present synthetic photometry simulated at different redshifts in the LSST filter system using the SN template built from the SN Ib iPTF13bvn.

These template SEDs can be easily integrated in SN simulation packages such as SuperNova ANAlysis software \citep[\snana;][]{2009PASP..121.1028K} or \textsc{sncosmo} \citep{2016ascl.soft11017B}. The Python package used to generate the templates is also available\footnote{\url{https://github.com/maria-vincenzi/PyCoCo}}, along with examples and tutorials.

\section{Data}
\label{sec:data}

We now describe the data from which we can use the techniques described in the previous section to construct a series of core collapse spectral templates. We do this using 67 core collapse SNe for which high-quality photometry and spectroscopy has been published. We describe the selection criteria we applied to build this dataset, and various other data preparation procedures.

\subsection{Description of the data set}
\label{sec:dataset_description}
\subsubsection{Selection criteria}
\label{sec:sn-selection}

We select core collapse SNe from the literature according to the following criteria:
\begin{itemize}
\item The published SNe must have photometric data in at least the three optical filters of $B$,$V$ and $R$ (or alternatively $g$,$r$ and $i$), with a well-constrained light curve interpolation and at least one epoch prior to peak brightness in the $V$ (or $g$) band. 
\item The SNe must have photometry in at least one near-UV filter, i.e., a filter that samples the wavelength range below 4000\AA\ in the SN rest-frame. This near-UV photometry must have at least one epoch within two days of the $V$/$g$ band peak.
\item The SNe must have at least five spectra, with at least one observation within three days of the $V$/$g$-band peak, and at least one observation after +15 days.
\end{itemize}
We do not apply any redshift constraints, and in principle our technique can be applied to SNe observed at any redshift. However, multiple spectroscopic follow-up observations of SNe become expensive at redshifts above $z\sim0.05$, and thus mostly nearby SN events have been included in this work.

We found 67 core collapse SN events in the literature that satisfy these quality cuts. This sample is presented in Table~\ref{summary_table:Phot_Spec_Ref}. Where available, we download the spectra from the WISeREP repository\footnote{\url{https://wiserep.weizmann.ac.il/}} \citep{2012PASP..124..668Y} and the photometry from the \lq Open Supernova Catalog\rq\ \citep{2017ApJ...835...64G}. 

\subsubsection{UV data}
\label{sec:swift}

Since all SNe selected are at low redshift ($z<0.03$; Table~\ref{summary_table:Phot_Spec_Ref}), near-UV photometry in the SN rest-frame corresponds to observations in the observer-frame $U$ band or bluer filters. For 36 of the 67 SNe selected in this work, UV photometry is measured by \textit{Swift}/UVOT  \citep{2014Ap&SS.354...89B} using four UV filters (and two optical filters) for imaging, and two grisms for low-resolution spectroscopy. The central wavelengths for the near-UV filters are 3465\AA, 2600\AA, 2246\AA\ and 1928\AA\ for the filters $u$, $uvw1$, $uvm2$, $uvw2$ respectively.

\subsubsection{Host galaxy extinction estimates}
\label{sec:host_extinction}

Providing a library of host galaxy extinction-corrected core collapse SN templates is a significant advantage, especially when using these templates in simulations, as arbitrary amounts of extinction can then be added. Thus the input to our template construction method in Section~\ref{sec:pycoco} should be extinction-corrected photometry.

However, estimating SN extinction due to local host galaxy dust is challenging as indicators of dust extinction are difficult to measure. We discuss various methods and resources used to estimate the SN host galaxy extinction in Appendix~\ref{Appendix:extinction}. In Table~\ref{summary_table:Phot_Spec_Ref} we report for each SN the values adopted for reddening due to dust in the SN host galaxy, $E(B-V)^\mathrm{host}$. We note that our code will make the required corrections to the SN photometry provided the extinction estimates (i.e., $E(B-V)$) are provided.

\subsubsection{Supernova classification scheme}
\label{sec:class_scheme}

In the application of our library of templates for simulations, and for training supervised machine learning classifiers, we require a broad classification scheme. Given the limited number of objects available in our dataset, we group our templates into six spectral sub-types: SN Ib (13 SNe), SN Ic (7 SNe), SN Ic-BL (6 SNe), SN IIb (11 SNe), SN II (23 SNe), SN IIn (7 SNe) and SN\,1987A.

We do not apply the historical distinction between SNe IIP and SNe IIL, as recent analyses show reduced evidence for such a separation \citep{2014ApJ...786...67A}. The sample of 23 SNe II considered in this work present a continuum of decline rates, from small decline rates and almost flat light curves, to fast-declining events fading at around one magnitude per 30 days. 

We also include in our sample more recently identified sub-classes of peculiar SNe, such as SNe Ibn (SN\,2010al) and other peculiar SN IIn events (IIn-pec) such as SN\,2009ip.
SNe Ibn/IIn-pec represent very interesting classes of rare transients:  their luminosities can be comparable to SNe Ia and their light curves are not necessarily as wide as normal SNe II. However, the lack of unbiased samples of these classes of transients, and their diversity, makes it difficult to measure their global properties (e.g., their luminosity distribution or relative rates), and therefore, we combine SNe Ibn and SNe IIn-pec into a single group of SNe IIn. In Table~\ref{summary_table:Phot_Spec_Ref}, we report the list of SNe included in this work and the spectroscopic type.

\subsection{Implementation}
\label{sec:data_prep}

The data that we use to build the spectral templates are generally used as published. However, in the implementation phase it is sometimes necessary to remove observational and astrophysical signatures that do not have a SN origin, or that are not relevant for the purpose of these templates. We also apply some basic data conversions and noise reduction procedures as follows:
\begin{enumerate}
    \item As the light curve interpolation is always performed in flux space, we convert photometric measurements from magnitudes to flux densities. For the conversion of photometry to flux densities we use our own custom-written software\footnote{\url{https://github.com/chrisfrohmaier/what_the_flux}}.
    
    \item{We correct the photometry for Milky Way and host galaxy extinction using a \cite{1989ApJ...345..245C} dust law with $R_V=3.1$. For Milky Way extinction we use the \citet{1998ApJ...500..525S} dust maps. We discuss in detail our choice of $R_V^{host}=3.1$ in Appendix \ref{Appendix:extinction}}.

    \item Some older events (e.g., SN\,1987A, SN\,1993J) do not have published uncertainties on their magnitude measurements. We arbitrarily estimate the uncertainties as 10 per cent of the fluxes.

    \item Photometric observations taken on the same night and in the same filter are averaged in order to reduce uncertainties in the individual measurements.
    
    \item Using a sigma-clipping algorithm, we remove spectral features clearly associated with a host galaxy (e.g., nebular emission lines) or due to the atmosphere (telluric features).

    \item We smooth each spectrum by applying a Savitzky-Golay filter \citep{1964AnaCh..36.1627S} using a window of 100\,\AA.

\end{enumerate}

Once the data has been pre-processed as described, we apply on them the technique presented in Section \ref{sec:pycoco}. In around 6 per cent of the light curves considered in our data set, the signal-to-noise of the photometry is sufficiently low such that the GPs over-smooth the data. In such cases, we manually fix the scale of the kernel or use a non-zero prior on the mean function when performing the light-curve fitting described in Section \ref{sec:light-curve-fits}.

\begin{figure*}
\centering
\includegraphics[width=0.9\textwidth]{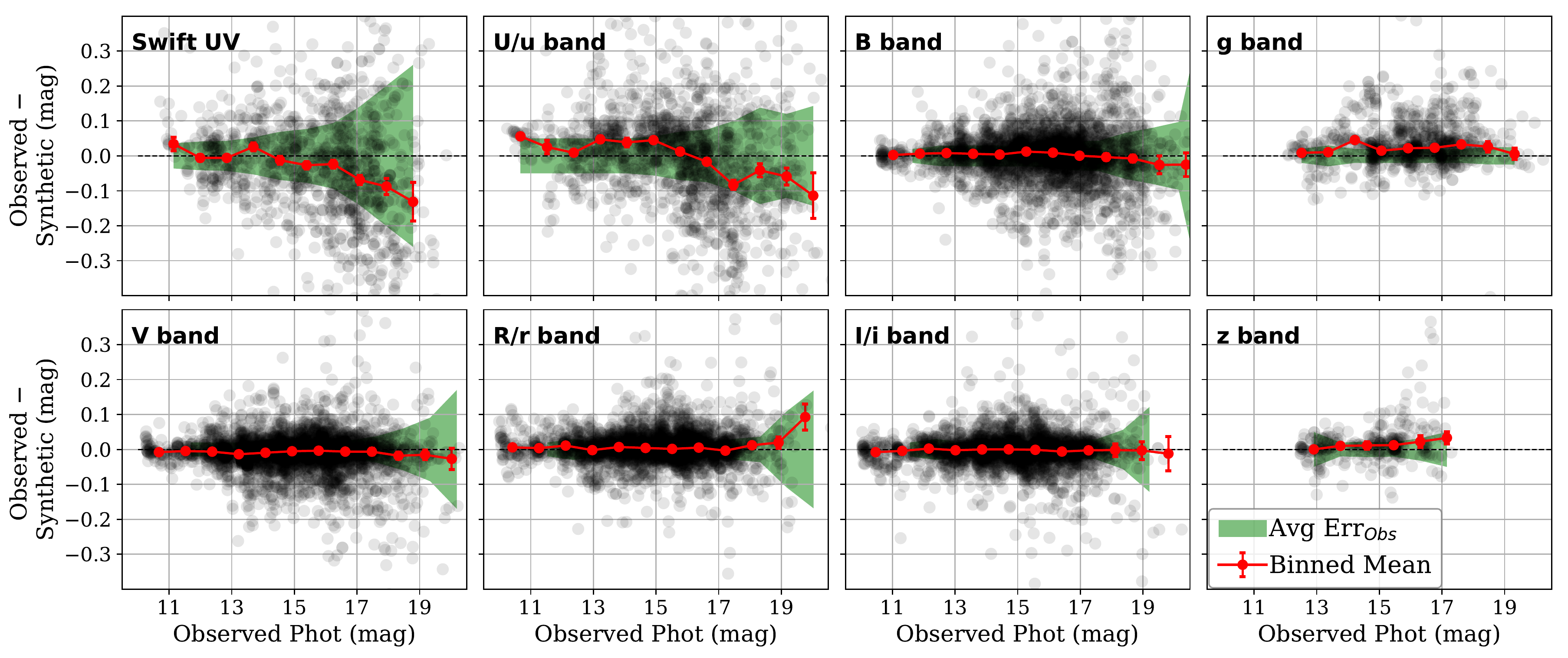}
\caption{Observed photometry compared to photometry synthesised from the final templates of the 67 SNe in our sample. The observed photometry is the published photometry corrected for Milky Way and host dust extinction, and it is the same used to perform the light curve fitting described in Section~\ref{sec:light-curve-fits}. Synthetic photometry at the same epoch of the observed photometry is measured from the final flux-calibrated, daily-sampled and near-UV extended spectral time-series. For every SN, for every photometric point observed for the SN, the difference between observed and synthetic photometry is shown (black symbols) as well as the binned mean and error on the mean of the residuals. Each panel shows the residuals for different group of filters: UV filters (\textit{Swift} filters $uvw2$, $uvm2$, $uvw1$), $u$ band (Bessell $U$, Swift $U$, sdss $u$ bands), $B$ band (Bessell $B$ and Swift $B$ bands), $V$ band (Bessell $V$ and Swift $V$ band), $g$ band, $R/r$ band (Bessell $R$ and sdss $r$ bands), $I/i$ band (Bessell $I$ and sdss $i$ bands), and $z$ band. Overall, we recover the observed photometry to within 0.02 mags in the optical bands and approximately 0.1 mags in the near-UV.}
\label{FIG:pycoco_residuals}
\end{figure*}

After testing our technique on the sample of 67 SNe described in Section \ref{sec:dataset_description}, we compare the published observed photometry with the photometry synthesised from the final flux-calibrated, daily-sampled and near-UV extended spectral time-series.
Our results are presented in Fig.~\ref{FIG:pycoco_residuals}. On average the observed photometry is recovered to within 0.02 magnitudes in most of the optical filters and to within 0.1 magnitudes in the near-UV. 

\section{Application to simulations of SN surveys}
\label{sec:applications}

Our new library of core collapse SN templates has many potential uses in photometric SN classification and the simulations of optical transient surveys. In this section, we demonstrate their application to SN simulations by integrating our library of core collapse SN templates into the SN light-curve simulation package \snana. This requires two further key ingredients in simulating SNe, no matter what the scientific goal: the luminosity function used to set the overall brightness distribution of the simulated events, and the relative rates that different SN subtypes are simulated with. We then present an example application: the simulation of contamination by core collapse SNe in cosmological surveys of SNe Ia.

\subsection{Luminosity functions and relative rates}
\label{sec:lf_rates}

The luminosity function (LF) describes the intrinsic distribution of brightnesses (absolute magnitudes) of SNe. Due to selection biases inherent in all astronomical surveys (or compilations/catalogues), it is difficult to measure accurately, with two main core collapse SN luminosity functions published in the last decade: \citet{2011MNRAS.412.1441L} and \citet{2014AJ....147..118R}.

The brightness distributions presented in \citet[][hereafter \citetalias{2011MNRAS.412.1441L}]{2011MNRAS.412.1441L} are measured from a volume-limited sample of 92 core collapse SNe from the Lick Observatory Supernova Search \citep[LOSS;][]{2011MNRAS.412.1419L}. They are constructed in the $R$-band filter, and calibrated to the \citet{1992AJ....104..340L} system from mostly unfiltered images with a precision of about five per cent \citep{2003PASP..115..844L}. No $k$-corrections were made as the SNe all lie within 60\,Mpc and thus  $k$-corrections are small. In order to correct for potential biases in the volume-limited sample, \citetalias{2011MNRAS.412.1441L} estimated for each SN the completeness of the LOSS survey within the fixed cutoff distance, given the brightness of the event. The peak absolute magnitude and light-curve shape of each SN are recovered by performing template fitting of the $R$-band light curve then used in the computation of the completeness \citep[see also][]{2011MNRAS.412.1419L}. Different templates are used for SNe of different types. Each SN in the LF sample is then weighted by the reciprocal of this completeness.

Recently, \citet{2017PASP..129e4201S} revised the classifications of the SNe in the LOSS sample, and presented an updated measurement of the relative rate of each SN sub-type. In particular, SNe IIL and IIP are now grouped into a single population (mirroring our templates), and calcium-rich SNe \citep{2010Natur.465..322P} and SNe Ic-BL are considered distinct classes. The classification of 15 SNe has also been corrected using additional spectroscopic observations. We use this re-classified LOSS sample to remeasure the LF of each core collapse SN sub-type, with SNe Ic, Ic-pec and Ic-BL grouped into the same sub-class due to the small numbers of events included in these classes (the relative fraction can be preserved in any simulations). We also exclude two SN IIn now reclassified as SN imposters.

In principle, for each re-classified SN a new template fit and the re-computation of the completeness following \citetalias{2011MNRAS.412.1441L} would be required. However, \citetalias{2011MNRAS.412.1441L} used the same template for SNe Ib and Ic, and thus SNe Ic re-classified as SNe Ib do not require new completeness calculations.

Other SNe that do require a new light curve fit and for which the completeness should be recalculated are partially or completely excluded. This is the case of 5 SNe. SN 2003bw, SN 2003br and SN 2005mg are re-classified from SNe II to SNe II/IIb. We can include them in SNe II LF but not in SNe IIb LF. SN 2004C and SN 2005lr are re-classified from Ic to IIb, therefore we can not include them in the SN IIb LF. These approximations are likely to bias the LF for SNe IIb. However, an accurate estimate of LFs goes beyond the purpose of this work. A comparison between the original \citetalias{2011MNRAS.412.1441L} LFs and revised ones is shown in Table~\ref{table:LFs}. 

We also consider the LFs of \citet[][hereafter \citetalias{2014AJ....147..118R}]{2014AJ....147..118R}. \citetalias{2014AJ....147..118R} measured LFs from a larger sample of 211 core collapse SNe, combining the Asiago SN Catalog \citep{2002AJ....123..745R,2014AJ....147..118R} and supplementary magnitude-limited SN samples. These LFs are presented in the $B$ band, and have additionally been corrected for host-galaxy extinction. However, only half of the SNe considered (113 of 211) have specific information about host galaxy extinction, with an average $B$-band extinction of 0.292\,mag assumed for the other events. Intrinsic biases in the SN sample are corrected using a bootstrap technique, and LFs measured from both the bias-corrected SN sample and from a volume-limited but \emph{not} bias-corrected SN sample. We test in our simulations the bias-corrected LFs (also reported in Table~\ref{table:LFs}).

To determine the relative fraction of each core collapse sub-type, we use the revised nearby SN relative rates presented in \citet[][see also Table~\ref{table:LFs}]{2017PASP..129e4201S}. These relative rates are assumed across all our simulations.

The LFs and the relative rates used in this section are all based on local SN measurements. In our simulations, we assume these LFs and relative rates also apply at higher redshift, i.e., we neglect any redshift evolution in these quantities. The only redshift dependency we assume is that of the global volumetric rate of all core collapse SNe.

\begin{table*}
\centering
    \caption{Summary of luminosity functions and relative rates for core collapse SNe}
    \label{table:LFs}
    \begin{tabular}{lccccc}
    \hline
      SN type&   LFs from & Revised LFs from &  LFs from & LFs adjusted from &Rates$^{d}$ \\
      &   \citet{2011MNRAS.412.1441L}$^{a}$ & \citet{2011MNRAS.412.1441L} $^{b}$ & \citet{2014AJ....147..118R}$^{c}$ & \citet{2017ApJ...843....6J} & \\
    \hline
    II &    - &  -15.97(1.31) & - & - & 64.9 \\
    IIL &   -17.44(0.64) &  - & -17.98(0.86)$^{f}$ & -18.28(0.45)& 7.9 \\
    IIP &   -15.66(1.23) &  - & -16.75(0.98)$^{f}$ & -16.67(1.08) & 57.0 \\
    IIb &   -16.65(1.30) &  -16.69(1.38) & -16.99(0.92) & -16.69(1.99)& 10.9 \\
    IIn &   -16.86(1.61) & -17.90(0.95) & -18.53(1.36) & -17.66(1.08)& 4.7 \\
    \hline
    Ic &    -16.04(1.28) &  -16.75(0.97) & - & -17.44(0.66) & 7.5 \\
    Ib &    -17.01(0.41) &  -16.07(1.34) & -17.45(1.12) & -18.26(0.15)& 10.8 \\
    Ibc-pec &    -15.50(1.21) &  - & - & - & - \\
    Ic/Ic-pec/Ic-BL  &  - &  -16.79(0.95) & -17.66(1.18) & - & 8.6 \\
    Ic-BL &     - &  - & - & - & 1.1 \\
    \hline
    \end{tabular}
    \begin{tablenotes}\footnotesize
        \item $^{a}$ Mean and r.m.s. of the $R$-band absolute magnitudes in the bias-corrected sample from \citet{2011MNRAS.412.1419L}. No host extinction correction.
        \item $^{b}$ Mean and r.m.s. of the $R$-band absolute magnitudes in the bias-corrected sample, using the \citet{2017PASP..129e4201S} classifications. No host extinction correction.
        \item $^{c}$ Mean and r.m.s. of the $B$-band absolute magnitudes in the bias-corrected sample from \citetalias{2014AJ....147..118R}. Host extinction corrections are applied.  
        \item $^{d}$ Relative SN rates from \cite{2017PASP..129e4201S}, with SN IIL and SN IIP rates taken from \citetalias{2011MNRAS.412.1441L}. Numbers are normalised to sum to one.
        \item $^{f}$ \citetalias{2014AJ....147..118R} considered IIL and IIP as separate classes. We adapt our library of templates to this classification scheme when simulating core collapse SNe using \citetalias{2014AJ....147..118R} luminosity functions, and we define SNe IIL/IIP as SNe with decline rates above/below 1.1 magnitudes per 100 days.
    \end{tablenotes}
\end{table*}

\subsection{Simulating supernovae in cosmological surveys}

SNe Ia can be used as standardisable candles to estimate distances in the universe \citep[][]{1998AJ....116.1009R,1999ApJ...517..565P}. Historically, samples of SNe Ia have been constructed where every SN Ia event has been both spectroscopically confirmed and has a spectroscopic redshift measurement \citep[e.g.,][]{2014AA...568A..22B,2018ApJ...859..101S}. However, contemporary and future surveys will be unable to spectroscopically confirm every SN Ia event detected due to a lack of follow-up resources, and will therefore be reliant on photometric classification techniques based on a multicolour light curve and possibly a spectroscopic redshift of the SN host galaxy.

Contamination from events that are not SNe Ia is therefore potentially a serious problem \citep{2013ApJ...763...88C, 2017ApJ...843....6J}. Simulations can be used to estimate selection effects and the level of this contamination \citep{2017ApJ...836...56K}; a recent example can be found in \citet[][hereafter \citetalias{2017ApJ...843....6J}]{2017ApJ...843....6J} and \citet[][]{2018ApJ...857...51J} using the the Pan-STARRS 1 (PS1) photometric SN sample.

These simulations have analysed the \lq Hubble residual\rq\ distribution for the SNe Ia in the sample. The Hubble residual is defined as $\mu_\mathrm{obs}-\mu_\mathrm{model}$, where $\mu_\mathrm{obs}$ is the observed distance modulus for each SN event, and $\mu_\mathrm{model}$ is the theoretical distance modulus expected for the redshift of each event given a set of cosmological parameters. $\mu_\mathrm{obs}$ is defined as
\begin{equation}
\mu_\mathrm{obs} = m_B + \alpha x_{1} - \beta \mathcal{C} - M_B,
\label{eqn:distmod}
\end{equation}
where $m_B$ is the rest-frame $B$-band apparent magnitude at maximum light (or more accurately the log of the light-curve amplitude in the light-curve fit), $x_1$ is the SN \lq stretch\rq\ parameter, $\mathcal{C}$ is the SN colour estimator, and $\alpha$, $\beta$ and $M_B$ are parameters defining the stretch--luminosity and colour--luminosity relationships for the SN Ia population being studied \citep[e.g.,][]{1998AA...331..815T,2006AA...447...31A}. The $x_1$ and $\mathcal{C}$ estimators are typically determined using the SALT2 light-curve fitter \citep{2007AA...466...11G}.

The set of simulations implemented for the cosmological analysis of the PS1 sample is presented in detail in \citetalias{2017ApJ...843....6J}. The parent distributions of $x_1$ and $\mathcal{C}$ used to simulate SNe Ia are determined by applying the method of \citet{2016ApJ...822L..35S} to the PS1 spectroscopic SN sample \citep{2014ApJ...795...44R} with additional adjustments, taking into account additional selection biases on the photometric sample. Core collapse SNe are simulated using the set of 41 templates from the SNPhotCC library, with the addition of four new templates of SNe IIb built using more recent observations (models for this SN sub-type were not included in the original SNPhotCC library). Templates of sub-luminous SN\,1991bg-like SNe Ia are also included. This library of templates is not corrected for host extinction and is matched to the original LFs taken from \citet{2011MNRAS.412.1441L}. This approach of modelling the core collapse SN population assumes that the reddening distribution in the data, the reddening distribution in the templates, and the reddening distribution in the sample of SNe used to measure the LFs are equal and representative.

A significant discrepancy is observed between the Hubble residual distribution
measured from the SN data, and that predicted from the simulations: a third of the SNe observed in the data with Hubble residuals in the range $0.5 < \mu_\mathrm{obs} - \mu_\mathrm{model} < 1.5$\,mag are not present in the simulations. In \citetalias{2017ApJ...843....6J}, this is accounted for by adjusting the core collapse LFs, brightening the \citet{2011MNRAS.412.1441L} values by $\simeq1$\,mag, and fine-tuning the width of the LF (these adjusted LFs are also reported in Table~\ref{table:LFs}). As discussed in \citetalias{2017ApJ...843....6J}, these adjustments are made after analysing the peak absolute brightnesses of the PS1 light curves identified as SNe Ib/Ic or SNe II by the photometric classifier \textsc{psnid} \citep{2011ApJ...738..162S}, after specific SALT2 cuts are applied to the sample, and are not necessarily representative of intrinsic SN LFs.

\subsubsection{Implementation of the simulations}
\label{sec:simulatons}

We repeat the \citetalias{2017ApJ...843....6J} simulations of the PS1 photometric SN sample using our new template library to demonstrate its applicability to simulations of SN surveys. Our simulations use the same \snana\ configuration files presented in \citetalias{2017ApJ...843....6J}, with the exception of revising the approach used to model and reproduce the core collapse population. We test three alternative ways to generate core collapse SNe:
\begin{enumerate}
    \item Using our uncorrected templates, matched with the revised \citetalias{2011MNRAS.412.1441L} LFs (hereafter referred to as V19+L11),
    \item Using our host-extinction corrected templates matched with the \citetalias{2014AJ....147..118R} LFs, simulating additional host extinction (\lq V19+R14 LFs\rq),
    \item Using our uncorrected templates, matched with the LFs adjusted by \citetalias{2017ApJ...843....6J} (\lq V19+J17\rq).
\end{enumerate}
For each approach we generate 25 independent simulations, and our results for each case are then an average of these 25.

The method used to match the templates to the relative LF is straight forward. For each of the SN sub-types considered in our classification scheme, we compare the (biased) luminosity distribution measured from our set of templates to the corresponding LF. Sub-type dependent magnitude offsets and smearing factors are applied to the templates so that the peak brightnesses of simulated SN events have the correct distribution. By applying \textit{sub-type dependent} magnitude offsets, we correct the overall brightness of each sub-class of templates, thus retaining the relative brightness between individual templates of the same sub-type. As a result, correlations intrinsic to sub-classes of templates (e.g., faster declining SNe II are on average brighter than slower declining SNe II) are preserved in the simulations.

The relative rate of each template is calculated so that the core collapse SN fractions presented in \citet{2017PASP..129e4201S} are correctly reproduced in the simulations. Within a sub-type, each individual template is considered to be equally probable. For simulations using templates corrected for host galaxy extinction, the effect of host galaxy dust is simulated using the dust extinction distribution presented in \citet*[][see Appendix~\ref{Appendix:extinction}]{1998ApJ...502..177H}.

We compare the results from these three approaches with the core collapse contamination modelling adopted in \citetalias{2017ApJ...843....6J} \textit{after} the fine-tuning of the LFs (hereafter referred as J17 templates with J17 adjusted LFs). This gives four families of simulations in total, three of which we run ourselves. The input files used to generate each simulation can be found in the package \snana.

\subsection{Results and Discussion}
\label{sec:sim-results}

We now turn to the results from our simulations. For each of the simulations, detected SNe are fitted using the SALT2.4 model \citep{2010AA...523A...7G}, and the  SALT2-based cuts of \citet{2014AA...568A..22B} and \citetalias{2017ApJ...843....6J} are then applied. In \citetalias{2017ApJ...843....6J}, over 25 independent simulations, the average number of simulated SNe Ia passing SALT2-based cuts is $1021 \pm 30$, which correspond to $88.4 \pm 3.7$ per cent of the number of SNe observed in the PS1 sample after SALT2-based cuts (1153 SNe $\pm$ a Poisson error of 34). Assuming that simulations of SNe Ia are correct and that each event in the data is either a SN Ia or a core collapse SN, we expect to reproduce with simulations a fraction of core collapse of $11.6 \pm 3.7$ per cent. 

\begin{figure*}
\centering
\includegraphics[width=0.95\textwidth]{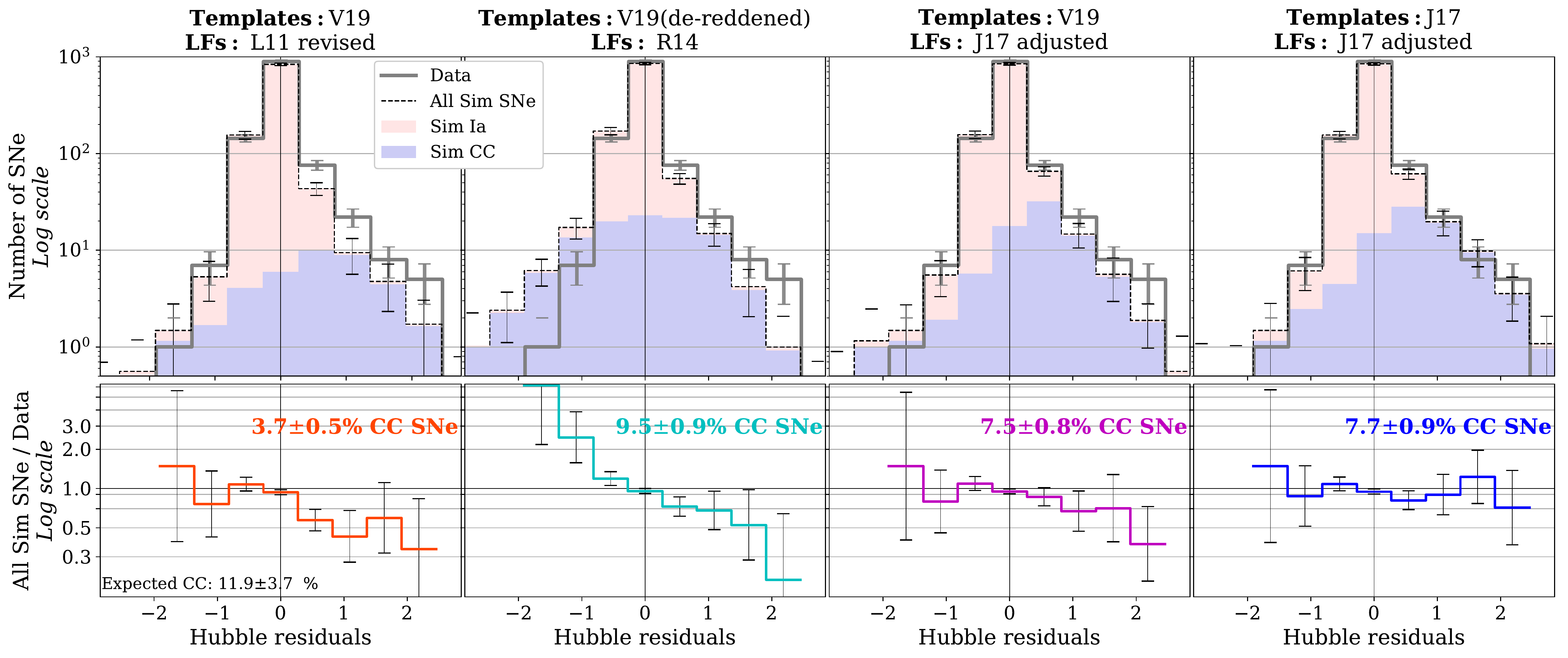}\\ 
\caption{\emph{Upper panels}: The Hubble residuals predicted i) using our library of core collapse templates generated without extinction corrections together with the revised \citetalias{2011MNRAS.412.1441L} luminosity functions (left), ii) using our library of extinction-corrected templates and the \citetalias{2014AJ....147..118R} luminosity functions (centre left), iii) using our library of core collapse templates generated without extinction corrections together with the luminosity functions adjusted by \citetalias{2017ApJ...843....6J}(centre right) and iv) using templates and luminosity functions adopted in \citetalias{2017ApJ...843....6J}.  Each panel shows the mean distribution of Hubble residuals averaged over 25 simulations. The distribution of Hubble residuals measured from all simulated SNe (core collapse SNe and SNe Ia) is shown in black. Uncertainties are measured as the standard deviation over 25 simulations. The red and blue histograms show the stacked distributions of simulated SNe Ia and simulated core collapse SNe respectively. The distribution of Hubble residuals measured from the PS1 data and the relative Poisson errors are shown in grey. 
\emph{Lower panels}: For each Hubble residual bin we present the ratio of the number of all simulated SNe (SNe Ia and core collapse) and the number of actual observed SNe with the relative propagated uncertainties. The fraction of simulated core collapse SNe over the total number of simulated SNe is indicated for each of the four simulations. The expected core collapse contamination (assuming the number of simulated SNe Ia is correct and events in the data are either SNe Ia or core collapse SNe) is also indicated. Except for simulations using luminosity functions and templates from \citetalias{2017ApJ...843....6J}, the simulations are not in good agreement with the data for Hubble residuals larger than 0.5 magnitude.}
\label{FIG:HR_PS1}
\end{figure*}

In Fig.~\ref{FIG:HR_PS1} we present for each simulation the predicted core collapse contamination and the distribution of Hubble residuals. (In the discussion that follows, we remind the reader that the Hubble residual is defined as $\mu_\mathrm{obs} - \mu_\mathrm{model}$; i.e. brighter SNe have a negative Hubble residual.)

Comparing our results to the PS1 data we observe that: 
\begin{itemize}
    \item Simulations performed using revised \citetalias{2011MNRAS.412.1441L} LFs and templates not corrected for host extinction (V19+L11) predict a core collapse contamination of $3.7 \pm 0.5$ per cent. The number of SNe in the fainter tail of the Hubble residuals is underestimated by a factor of approximately two. This reproduces the result of \citetalias{2017ApJ...843....6J} before the fine-tuning of the original \citetalias{2011MNRAS.412.1441L} LFs, and shows that \citetalias{2011MNRAS.412.1441L} LFs, used either in their original or revised form, underestimate the number of observed contaminants.
    \item Simulations performed using \citetalias{2014AJ....147..118R} LFs and de-reddened templates not only underestimate the number of SNe in the fainter tail of the Hubble residuals but also overestimate the number of bright events (Hubble residuals with $<-1$\,mag) by a factor of three. The overall predicted contamination is $9.5\pm0.9$ per cent. The high fraction of bright events suggest that either the \citetalias{2014AJ....147..118R} LFs are over-represented in bright objects, or the dust extinction applied in the simulations is underestimated (globally or for particular SN sub-types).
    \item Simulations using the LFs adjusted by \citetalias{2017ApJ...843....6J} (V19+J17 and J17 simulations) better reproduce the overall Hubble residual distribution as expected, and predict a contamination of $7.5\pm0.8$ and $7.7\pm0.9$ per cent respectively. However, we note that when our library of templates is used instead of the original \citetalias{2017ApJ...843....6J} library, the agreement between the simulations and data slightly decrease. This suggests that LFs fine-tuned by \citetalias{2017ApJ...843....6J} should be used with caution as they are fine-tuned to a specific set of core collapse templates.
\end{itemize}

\begin{figure*}
\centering
\includegraphics[width=0.9\textwidth]{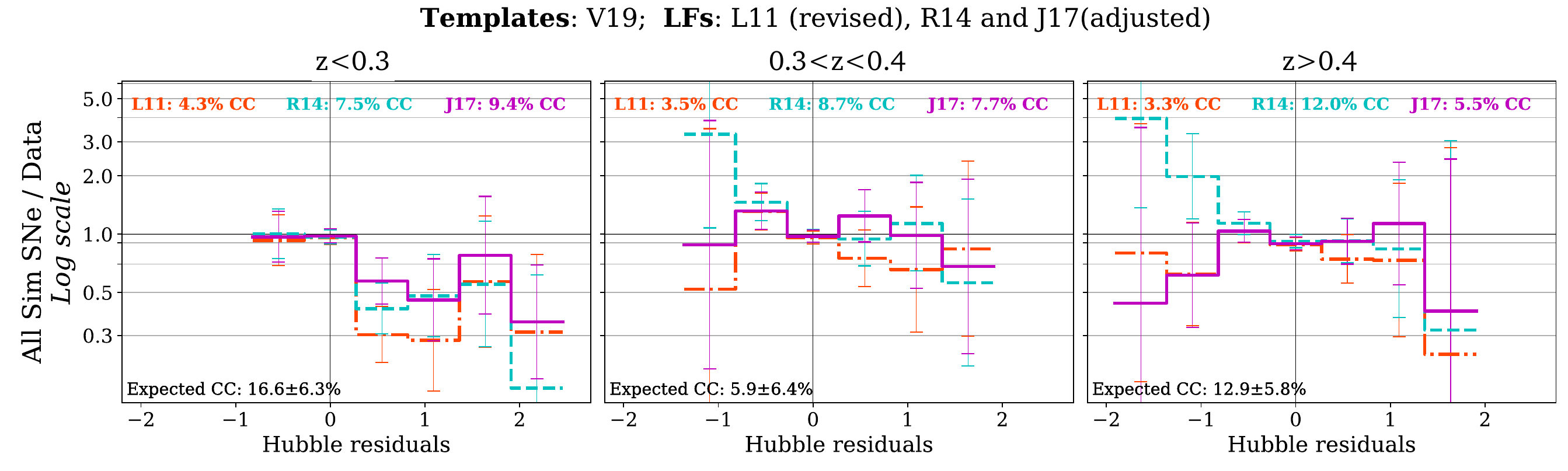}
\includegraphics[width=0.9\textwidth]{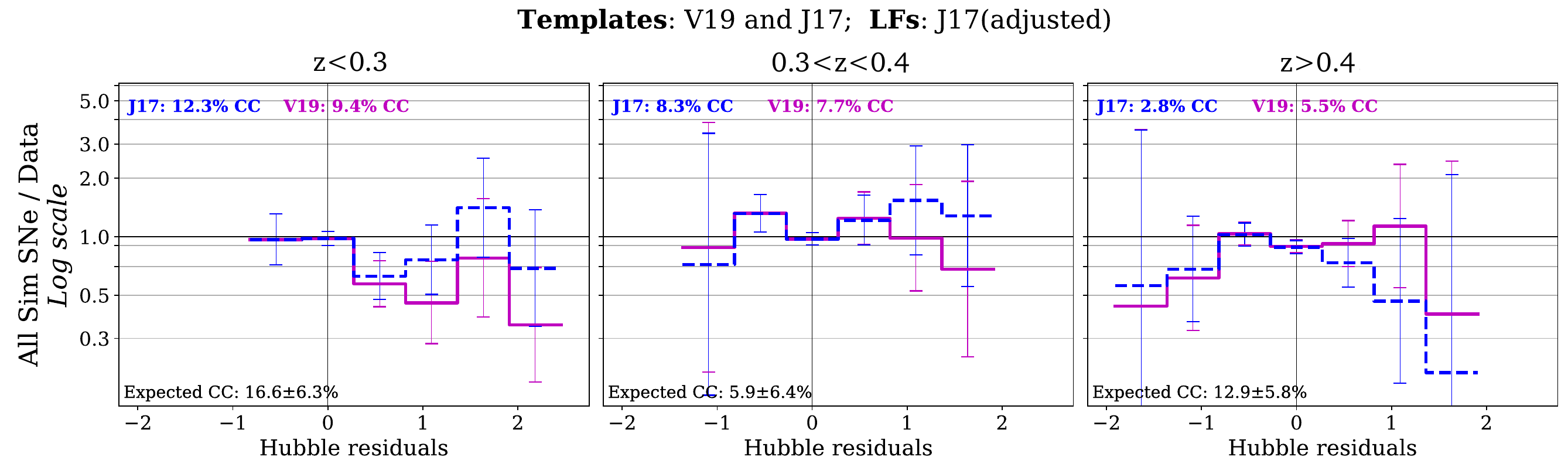}
\caption{Ratio between the number of simulated SNe and number of observed SNe in each Hubble residual bins for three different redshift intervals: $z<0.3$ (left, 368 SNe observed), $0.3<z<0.4$ (centre, 368 SNe observed), and $z>0.4$ (right, 417 SNe observed). \emph{Upper figure}: We compare the three simulations implemented using our library of templates, combined with three different LFs: revised \citetalias{2011MNRAS.412.1441L} LFs (dash-dotted line), \citetalias{2014AJ....147..118R} LFs (dashed line) and \citetalias{2017ApJ...843....6J} fine-tuned LFs (solid line). \emph{Lower plot}: We compare the two simulations implemented using LFs fine tuned by \citetalias{2017ApJ...843....6J} and two different template libraries: the template library presented in this work (V19, solid line) and the \citetalias{2017ApJ...843....6J} set of templates (dashed line). In every panel the percentage of simulated core collapse SNe over the total number of simulated SNe is shown for each simulation in each redshift bin. The expected core collapse contamination (assuming the number of simulated SNe Ia is correct and events in the data are either SNe Ia or core collapse SNe) is also indicated.}
\label{FIG:HR_redshift_bins}
\end{figure*}

In Fig.~\ref{FIG:HR_redshift_bins} we compare simulated and observed Hubble residual distributions in different redshift bins ($z<0.3$, $0.3<z<0.4$ and $z>0.4$). We present separately the effects of the different LFs while fixing the template library (upper plot in Fig.~\ref{FIG:HR_redshift_bins}) and then compare simulations implemented using the same LFs (\citetalias{2017ApJ...843....6J} adjusted) but different templates libraries (lower plot in Fig.~\ref{FIG:HR_redshift_bins}). Different redshift ranges present different issues:
\begin{itemize}
    \item[\textendash]{At lower redshifts ($z<0.3$) none of the simulations generated using our templates agree with the data. Despite the fact that the LFs tested are significantly different in terms of brightness (see Table~\ref{table:LFs}), the number of SNe for Hubble residuals $>0.5$\,mag is systematically under-predicted by a factor of two to three (corresponding to 20-25 \lq missing\rq\ SNe). Even using de-reddened templates and simulating a range of dust extinctions (V19+R14 set of simulations), we do not reproduce the observed contamination despite increasing significantly the colour diversity of the simulated light-curves. 
    At low redshifts, fainter SN events (e.g., stripped-envelope SNe or extincted SNe) are more likely to be detected; therefore the diversity of the template library, the fraction of peculiar events included, and the extinction distribution assumed for each SN sub-type, all have a significant impact on the final results.}
 
    \item[\textendash]{At higher redshift the fraction of detected core collapse SNe is reduced and the agreement between the data and simulations improves significantly. At $z>0.4$, we note that the expected core collapse fraction is larger than $12$ per cent (with $366\pm16$ simulated SNe Ia in this redshift bin, and 417 observed SNe in the PS1 sample). This raises some concerns about the modelling of SNe Ia in this redshift bin. We note that simulations using the \citetalias{2014AJ....147..118R} LFs can predict such a high contamination, but significantly over-predict the number of bright SNe (Hubble residuals of $<-1$\,mag). Simulations using the \citetalias{2017ApJ...843....6J} LFs better reproduce the Hubble residuals in this redshift bin, but here we note that when our \lq near-UV extended\rq\ templates are used, the predicted core collapse contamination is two times larger than that produced with other templates. This suggests that the near-UV extension is important for simulations of core collapse contamination at higher redshift.}

\end{itemize}

\begin{figure*}
\centering
\includegraphics[width=0.95\textwidth]{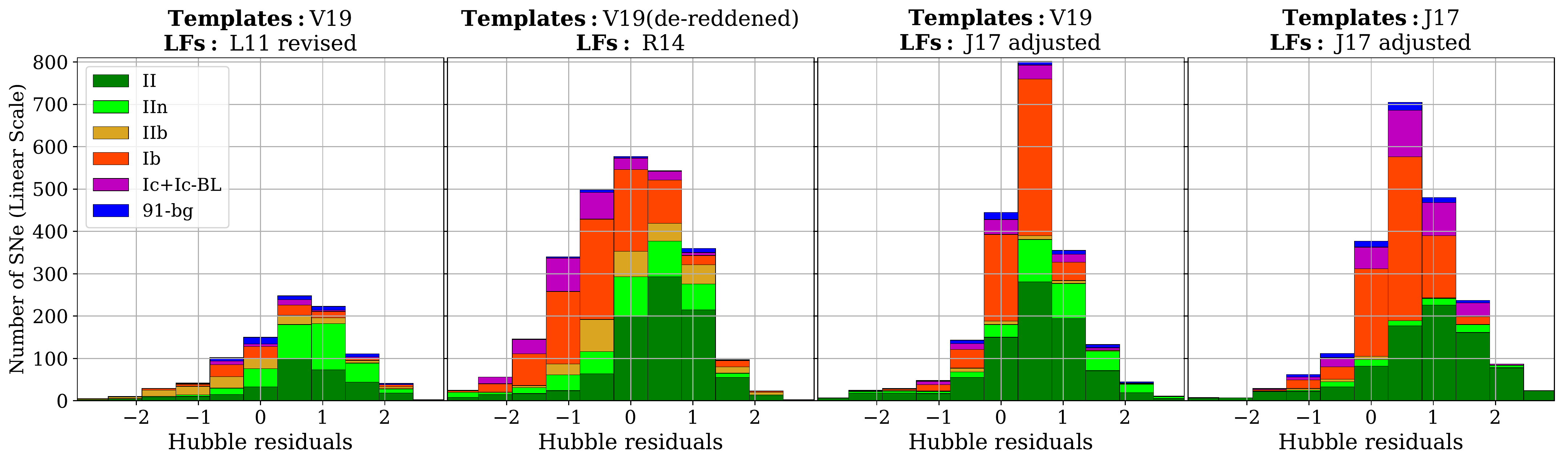}
\caption{Stacked distributions of simulated Hubble residuals for different core collapse SN types showing the relative importance of each SN type as a contaminant. The four panels show the results for the three simulations tested in this paper (Section~\ref{sec:simulatons}) plus the \citetalias{2017ApJ...843....6J} simulations. The results are the average of 25 different simulations.}
\label{FIG:HR_type_dependent}
\end{figure*}

Finally, Fig.~\ref{FIG:HR_type_dependent} shows the contribution of each core collapse SN type to the final simulated Hubble residual distributions. Several interesting trends emerge, highlighting the importance of the template library used in the simulations. For example, the number of SNe Ib and SNe Ic with Hubble residuals between $-1$ and $+1$\,mag is larger for simulations in which the \citetalias{2014AJ....147..118R} LFs and de-reddened templates are used. Two effects drives this result. Firstly, a significant fraction of the SNe Ib and Ic entering our templates are highly reddened ($E(B-V)>0.2$; see Fig.~\ref{FIG:reddening}), and many simulated SNe Ib and Ic do not pass the SALT2 colour cuts if templates that include host extinction are used. Second, the \citetalias{2014AJ....147..118R} LFs for SNe Ib ans SNe Ic are brighter than that of  \citetalias{2011MNRAS.412.1441L} or \citetalias{2017ApJ...843....6J}, even accounting for the differences in filters in which the two functions are constructed ($B$ band versus $R$ band).

We also note that more SNe IIn are detected in simulations using our templates, even when the same LF is used (third and fourth panel in Fig.~\ref{FIG:HR_type_dependent}). This is due to the larger diversity of SNe IIn captured in our library of templates, with both fast- and slow-declining SN IIn events included. In simulations where the L11 revised LFs and \citetalias{2014AJ....147..118R} LFs are used, a larger number of SNe IIn is expected as the intrinsic brightness of SNe IIn is larger than in the adjusted LFs of \citetalias{2017ApJ...843....6J} (see Table~\ref{table:LFs}). Finally, we observe that the Hubble residual distribution of SNe II from simulations using our library of templates is shifted by approximately 0.5 magnitudes when compared to simulations using the \citetalias{2017ApJ...843....6J} templates. This is true even when the same underlying LFs are used (third and fourth panel in Fig.~\ref{FIG:HR_type_dependent}) and is related to differences in the colour and stretch properties of the templates.

 
To conclude, analysing the Hubble residual distribution (globally, for different redshift bins and for different SN sub-types) is an important test for cosmological analyses. 
However, Hubble residuals are (by definition) conflated with SN parameters such as stretch and colour (equation~\ref{eqn:distmod}), and various cuts based on these SN properties are also applied.  
Thus, the analysis of Hubble residuals alone cannot provide a fundamental understanding of whether it is the luminosity functions, or the colour and stretch distributions and associated cuts, or both, that are driving the differences between observed and simulated samples.
We leave for a future article a detailed comparison between various global properties of observed and simulated SNe, using the samples both before and after SN parameter based cuts. 
However, understanding what drives the discrepancies between the data and simulations is crucial and provides important guidance in the design of realistic simulations of SNe of multiple types, essential to reduce systematic uncertainties due to core collapse contamination in future SN Ia analyses.

\section{Summary and future applications}\label{sec:summary}

We have presented a library of 67 spectral time-series templates of core collapse SNe. Compared to existing core-collapse spectral models, our new set of templates represent a significant improvement in several ways. We highlight the following key features of our library:
\begin{enumerate}
\item Spectral templates included in the library are generated from multi-band photometry and only sparsely-sampled spectral data.
\item Each template is generated from a single SN event, retaining the event-to-event diversity in the final library. No SED template or model is assumed, and the final daily-sampled spectral time series for each SN is reconstructed in a fully data-driven fashion, with a heavy emphasis on Gaussian processes. 
\item Each template is extended in the near-UV to $\simeq$1600\AA\ using near-UV photometry. This is critical to simulate SN light-curves at redshifts greater than 0.2. 
\item Each template has (optionally) been corrected for dust extinction in the SN host galaxy. As a result, different levels of host galaxy extinction, or different dust models, can be applied to the templates in simulations, and a realistic range of reddened SN light-curves can be reproduced. 
\end{enumerate}
The techniques and code used to construct the spectral library are open-source, and can, in principle, be applied to any type of transient with well-constrained multi-band photometry and spectroscopy. Given the large number of ongoing photometric and spectroscopic SN surveys, the number of SN events suitable for our method is likely to grow. This will boost our ability to improve and increase the diversity and size of core collapse SN libraries, of key importance to future SN Ia cosmological studies.

As a demonstration of the spectral library, we used the core collapse templates to simulate core collapse contamination in photometrically-selected SN Ia samples. We tested different core collapse modelling approaches, exploiting the luminosity functions and host extinction distributions currently available in the literature. Our analysis suggests that predictions from simulations are  sensitive to how global core collapse SN properties (such as luminosity, colour and extinction distributions) are modelled. This suggests that, along with these improved libraries of core collapse SN templates, it is critical to improve our knowledge of the global populations of core collapse SN events. 

\section*{Acknowledgements}
We thank Rick Kessler and Dan Scolnic for their useful comments and feedback. This work was supported by the Science \& Technology Facilities Council (STFC) (grant number ST/P006760/1) through the DISCnet Centre for Doctoral Training. MS acknowledges funding from the STFC (grant number ST/N002504/1) and from EU/FP7-ERC grant no. [615929]. This work made use of WISeREP (\url{https://wiserep.weizmann.ac.il}) and the Open Supernova Catalog (\url{https://sne.space}). This research has made use of the CfA Supernova Archive, which is funded in part by the National Science Foundation through grant AST 0907903. This research made use of Astropy,\footnote{http://www.astropy.org} a community-developed core Python package for Astronomy \citep{astropy:2013, astropy:2018}.

\bibliographystyle{mnras}
\bibliography{Biblio}

\begin{table*}
   \caption{The core collapse SNe included in this work.}
 \centering
 \label{summary_table:Phot_Spec_Ref}
\begin{tabular}{llllllcclc}
\hline
      Name &     Type & Redshift &  Peak M$_B$  & Optical data &    Near-UV data &  Numb. of  &      Ref.$^{(c)}$ &$E(B-V)_{host}^{(a)}$ & Ref.\\
   &       &  &   &  &  &  Spectra  &  & & Host$^{(c)}$\\
\hline
SN2013by &     II &  0.00359 & -18.36 &         $BVugri$ &    u,v,uvw1,uvm2,uvw2 &        7 &       (68),(69),(42) &              0 &     (69) \\
 ASASSN15oz &     II &  0.00693 & -18.19 &       $UBVRIgri$ &  u,b,v,uvw1,uvm2,uvw2 &       13 &                  (2) &              0 &      (2) \\
    SN2014G &     II &  0.00563 & -18.10 &          $UBVRI$ &      v,uvw1,uvm2,uvw2 &       17 &       (77),(71),(78) &           0.24 &     (78) \\
   SN2007od &     II &  0.00586 & -17.93 &        $UBVRIri$ &  u,b,v,uvw1,uvm2,uvw2 &       13 &                 (29) &              0 &     (29) \\
   SN2013ej &     II &  0.00219 & -17.76 &      $UBVRIgriz$ &  u,b,v,uvw1,uvm2,uvw2 &       28 &                 (73) &          0.049 &     (94) \\
   SN2013fs &     II &  0.01185 & -17.73 &          $BVRIr$ &      u,uvw1,uvm2,uvw2 &       23 &  (74),(63),(42),(75) &          0.015 &     (75) \\
   SN2008bj &     II &    0.019 & -17.48 &          $UBVri$ &                       &       16 &                 (31) &      0.081\dag &     (90) \\
    SN2016X &     II &  0.00441 & -17.38 &       $UBVRIgri$ &  u,b,v,uvw1,uvm2,uvw2 &       33 &                 (79) &           0.02 &     (79) \\
   SN2009bw &     II &  0.00382 & -17.14 &          $UBVRI$ &             uvm2,uvw2 &       16 &                 (39) &           0.08 &     (39) \\
   SN2013ab &     II &  0.00532 & -17.14 &        $BVRIgri$ &    u,b,uvw1,uvm2,uvw2 &       22 &            (64),(65) &           0.02 &     (65) \\
   SN2004et &     II &  0.00105 & -17.09 &          $UBVRI$ &                       &       28 &                 (16) &           0.07 &     (87) \\
   SN2012aw &     II &   0.0026 & -17.05 &     $UBVRIugriz$ &  u,b,v,uvw1,uvm2,uvw2 &       30 &       (61),(62),(63) &          0.028 &     (62) \\
   SN2008in &     II &  0.00522 & -16.96 &           $BVri$ &  u,b,v,uvw1,uvm2,uvw2 &       10 &            (36),(31) &          0.076 &     (36) \\
   SN2009dd &     II &  0.00246 & -16.81 &        $UBVRIri$ &                       &        6 &            (30),(31) &          0.433 &     (30) \\
   SN1999em &     II &  0.00239 & -16.69 &          $UBVRI$ &                       &       27 &                 (12) &          0.077 &     (86) \\
   SN2009kr &     II &   0.0065 & -16.57 &           $BVRI$ &  u,b,v,uvw1,uvm2,uvw2 &        6 &                 (44) &              0 &     (44) \\
 ASASSN14jb &     II &  0.00603 & -16.11 &          $BVgri$ &    u,b,uvw1,uvm2,uvw2 &       10 &                  (1) &              0 &      (1) \\
    SN2012A &     II &  0.00231 & -16.10 &          $UBVRI$ &                       &       39 &                 (57) &          0.012 &     (57) \\
   SN2009ib &     II &  0.00435 & -15.86 &      $BVRIugriz$ &                       &        9 &                 (40) &          0.131 &     (40) \\
   SN2013am &     II &  0.00269 & -15.80 &         $UBVRIi$ &    u,b,uvw1,uvm2,uvw2 &       14 &            (66),(67) &           0.55 &     (67) \\
    SN2009N &     II &  0.00345 & -15.03 &        $BVRgriz$ &  u,b,v,uvw1,uvm2,uvw2 &       22 &                 (37) &          0.113 &     (37) \\
   SN2005cs &     II &  0.00154 & -14.90 &         $UBVRIz$ &      b,uvw1,uvm2,uvw2 &       18 &                 (23) &          0.015 &     (89) \\
  SN2016bkv &     II &    0.002 & -14.89 &       $UBVRIgri$ &                       &       15 &            (80),(81) &              0 &     (81) \\
    SN1987A &     \tiny{87A-like} &  0.00001 & -14.33 &          $UBVRI$ &                       &       18 &                  (3) &           0.17 &     (85) \\
   \hline
   SN2011fu &    IIb &   0.0185 & -18.62 &         $UBVRIz$ &                       &       28 &       (52),(53),(21) &          0.015 &     (53) \\
    SN2006T &    IIb &   0.0081 & -17.72 &         $BVugri$ &                       &        6 &        (8),(19),(18) &  0.277 (0.048)$^{(b)}$ &     (88) \\
   SN2008ax &    IIb &   0.0019 & -16.65 &     $UBVRIugriz$ &       u,b,v,uvw1,uvw2 &       42 &        (34),(8),(24) &           0.28 &     (91) \\
  SN2016gkg &    IIb &  0.00492 & -16.61 &         $BVRIgr$ &  u,b,v,uvw1,uvm2,uvw2 &       17 &                 (82) &           0.09 &     (95) \\
   SN2008aq &    IIb &  0.00797 & -16.60 &         $BVugri$ &       u,b,v,uvw1,uvw2 &        7 &             (8),(18) &  0 (0)$^{(b)}$ &     (88) \\
   SN2011ei &    IIb &   0.0093 & -16.60 &           $BVRI$ &       u,b,v,uvw1,uvw2 &       13 &            (50),(51) &           0.18 &     (51) \\
   SN2011dh &    IIb &   0.0015 & -15.99 &         $UBVRIg$ &                       &       39 &            (48),(49) &           0.05 &     (49) \\
   SN2013df &    IIb &  0.00239 & -15.90 &          $UBVRI$ &  u,b,v,uvw1,uvm2,uvw2 &       12 &  (70),(71),(72),(21) &           0.08 &     (72) \\
   SN2011hs &    IIb &  0.00571 & -15.71 &      $BVRIugriz$ &       u,b,v,uvw1,uvw2 &       13 &                 (54) &           0.16 &     (54) \\
   SN2008bo &    IIb &    0.005 & -15.46 &           $BVri$ &           u,uvw1,uvw2 &       13 &        (8),(24),(21) &          0.033 &     (90) \\
    SN1993J &    IIb &  -0.0001 & -13.22 &          $UBVRI$ &                       &       50 &              (4),(5) &            0.1 &      (5) \\
    \hline
   SN2008fq &    IIn &   0.0106 & -18.89 &         $BVugri$ &                       &       10 &            (35),(25) &           0.46 &     (25) \\
   SN2010al &    IIn &    0.017 & -18.75 &        $BVRIuri$ &  u,b,v,uvw1,uvm2,uvw2 &       11 &  (45),(46),(31),(21) &          0.053 &     (46) \\
   SN2007pk &    IIn &  0.01665 & -18.71 &           $BVri$ &  u,b,v,uvw1,uvm2,uvw2 &        9 &            (30),(31) &      0.081\dag &     (91) \\
   SN2009ip &    IIn &  0.00594 & -18.33 &           $BVRI$ &      u,uvw1,uvm2,uvw2 &       55 &            (41),(42) &           0.01 &     (91) \\
   SN2006aa &    IIn &   0.0207 & -17.61 &         $BVugri$ &                       &        6 &                 (25) &      0.081\dag &     (90) \\
   SN2011ht &    IIn &   0.0036 & -16.98 &           $BVRI$ &  u,b,v,uvw1,uvm2,uvw2 &       12 &            (55),(56) &          0.061 &     (56) \\
   \hline
   SN2005bf &     Ib &   0.0189 & -18.44 &         $BVugri$ &                       &       26 &   (22),(8),(18),(21) &            0.1 &     (22) \\
   SN2007uy &     Ib &    0.007 & -18.34 &           $BVri$ &  u,b,v,uvw1,uvm2,uvw2 &       11 &        (8),(24),(21) &           0.63 &     (93) \\
   SN2005hg &     Ib &    0.021 & -18.02 &          $UBVri$ &                       &       20 &        (8),(24),(21) &     0.2235\dag &     (90) \\
   SN2009iz &     Ib &   0.0142 & -17.69 &          $BVuri$ &                  uvw1 &       10 &        (8),(24),(21) &     0.2235\dag &     (90) \\
   SN2009jf &     Ib &    0.008 & -17.68 &  $UBVRIuugrriiz$ &                       &       31 &   (43),(8),(24),(21) &           0.03 &     (91) \\
   SN2004gv &     Ib &     0.02 & -17.39 &         $BVugri$ &                       &        5 &        (8),(18),(21) &  0.053 (0.118)$^{(b)}$ &     (88) \\
   SN2012au &     Ib &   0.0045 & -17.39 &          $BVRIz$ &  u,b,v,uvw1,uvm2,uvw2 &        5 &       (59),(60),(21) &              0 &     (60) \\
   SN2006ep &     Ib &    0.015 & -17.16 &         $BVugri$ &                       &        7 &        (8),(18),(21) &  0.233 (0.226)$^{(b)}$ &     (88) \\
   SN2004gq &     Ib &  0.00647 & -16.97 &         $BVugri$ &                       &       24 &        (8),(19),(18) &  0.067 (0.143)$^{(b)}$ &     (88) \\
   SN1999dn &     Ib &   0.0093 & -16.62 &          $UBVRI$ &                       &       13 &                 (11) &          0.048 &     (11) \\
  iPTF13bvn &     Ib &  0.00449 & -16.54 &      $UBVRIgriz$ &       u,b,v,uvw1,uvw2 &       25 &            (83),(84) &           0.17 &     (84) \\
    SN2008D &     Ib &    0.007 & -16.47 &           $BVri$ &              u,v,uvw1 &       20 &   (33),(8),(24),(21) &            0.6 &     (91) \\
    SN2007Y &     Ib &   0.0046 & -16.32 &         $BVugri$ &  u,b,v,uvw1,uvm2,uvw2 &        6 &       (27),(18),(21) &  0.090 (0.318)$^{(b)}$ &     (27) \\
  \hline
\end{tabular}
\end{table*}

\begin{table*}
   \contcaption{}
\begin{tabular}{llllllcclc}
\hline
      Name &     Type & Redshift &  Peak M$_B$  & Optical data &    Near-UV data &  Numb. of  &      Ref.$^{(c)}$ &$E(B-V)_{host}^{(a)}$ & Ref.\\
   &       &  &   &  &  &  Spectra  &  & & Host$^{(c)}$\\
\hline
   SN2004aw &     Ic &   0.0159 & -17.66 &          $UBVRI$ &                       &       23 &                 (15) &           0.35 &     (15) \\
    SN1994I &     Ic &   0.0015 & -17.05 &          $UBVRI$ &                       &       27 &          (6),(7),(8) &           0.42 &      (7) \\
   SN2004fe &     Ic &    0.018 & -17.05 &         $BVugri$ &                       &        9 &        (17),(8),(18) &  0.000 (0.052)$^{(b)}$ &     (88) \\
   SN2004gt &     Ic &  0.00548 & -16.66 &         $BVugri$ &                       &       13 &   (20),(8),(18),(21) &          0.237 &     (88) \\
   SN2013ge &     Ic &   0.0055 & -16.57 &          $VRIri$ &  u,b,v,uvw1,uvm2,uvw2 &       18 &                 (76) &          0.047 &     (76) \\
   SN2007gr &     Ic &   0.0017 & -16.20 &        $UBVRIri$ &                       &       36 &   (28),(8),(24),(21) &           0.03 &     (92) \\
   SN2011bm &     Ic &   0.0022 & -13.10 &     $UBVRIugriz$ &                       &       13 &                 (47) &          0.032 &     (47) \\

   \hline
   SN2006aj &  Ic-BL &   0.0334 & -19.33 &          $UBVri$ &  u,b,v,uvw1,uvm2,uvw2 &       22 &        (26),(8),(24) &            0.2 &     (91) \\
   SN1998bw &  Ic-BL &   0.0085 & -18.93 &          $UBVRI$ &                       &       21 &             (9),(10) &              0 &     (10) \\
   SN2009bb &  Ic-BL &     0.01 & -18.72 &         $BVugri$ &                       &       12 &            (38),(18) &  0.540 (0.200)$^{(b)}$ &     (88) \\
   SN2007ru &  Ic-BL &   0.0155 & -18.65 &          $UBVRI$ &                       &        5 &        (32),(8),(21) &              0 &     (32) \\
   SN2012ap &  Ic-BL &   0.0121 & -17.66 &           $BVRI$ &            u,b,v,uvw1 &       12 &            (58),(21) &            0.4 &     (58) \\
   SN2002ap &  Ic-BL &   0.0022 & -17.07 &          $UBVRI$ &                       &       23 &        (13),(14),(8) &          0.008 &     (14) \\
\end{tabular}
    \begin{tablenotes}\footnotesize
    \item $^{(a)}$ Source of the estimate is given in the Reference column. Unless otherwise indicated, reddening estimates presented in the Table are derived from measurements of the equivalent width (EW) of the \ion{Na}{i}\,D absorption lines, and applying an empirical relation $E(B-V)^{host}$ = 0.16 EW$_{\ion{Na}{i}\,D}$ \citep{2003fthp.conf..200T}
    \item $^{(b)}$ Reddening inferred from colour evolution using the method presented in \citet{2018AA...609A.135S}. Estimates using the \ion{Na}{i}\,D absorption lines are shown in parenthesis for comparison.
    \item $\dag$ Estimate of host reddening not available. For this object the median extinction estimated in \cite{2016MNRAS.458.2973P} is used instead.
    \item $^{(c)}$ \emph{References:} (1) \citet{2018arXiv181111771M}; (2) \citet{2019arXiv190109962A}; (3) \citet{1987MNRAS.229P..15C}  \citet{1988MNRAS.231P..75C}  \citet{1988MNRAS.234P...5W}  \citet{1989MNRAS.237P..55C}  \citet{1995ApJS...99..223P}; (4) \citet{1995AAS..110..513B}  \citet{2000AJ....120.1487M}  \citet{2000AJ....120.1499M}; (5) \citet{1994AJ....107.1022R}; (6) \citet{1995ApJ...450L..11F}  \citet{1996ApJ...462..462C}; (7) \citet{1996AJ....111..327R}; (8) \citet{2014AJ....147...99M}; (9) \citet{1998Natur.395..670G}; (10) \citet{2001ApJ...555..900P}; (11) \citet{2011MNRAS.411.2726B}; (12) \citet{2001ApJ...558..615H}  \citet{2002PASP..114...35L}  \citet{2016AJ....151...33G}; (13) \citet{2002MNRAS.332L..73G}  \citet{2003PASP..115.1220F}  \citet{2013ApJ...767..162C}; (14) \citet{2003MNRAS.340..375P}; (15) \citet{2006MNRAS.371.1459T}; (16) \citet{2006MNRAS.372.1315S}  \citet{2010MNRAS.404..981M}  \citet{2014MNRAS.442..844F}; (17) \citet{2008AA...488..383H}; (18) \citet{2018AA...609A.134S}; (19) \citet{2017PASP..129e4201S}; (20) \citet{2008ApJ...687L...9M}  \citet{2009MNRAS.397..677T}; (21) \citet{2019MNRAS.482.1545S}; (22) \citet{2006ApJ...641.1039F}; (23) \citet{2006MNRAS.370.1752P}  \citet{2009ApJ...700.1456B}  \citet{2009MNRAS.394.2266P}; (24) \citet{2014ApJS..213...19B}; (25) \citet{2013AA...555A..10T}; (26) \citet{2006ApJ...645L..21M}  \citet{2006Natur.442.1011P}  \citet{2008AstBu..63..228S}; (27) \citet{2009ApJ...696..713S}; (28) \citet{2008ApJ...673L.155V}  \citet{2014ApJ...790..120C}; (29) \citet{2011MNRAS.417..261I}; (30) \citet{2013AA...555A.142I}; (31) \citet{2017ApJS..233....6H}; (32) \citet{2009ApJ...697..676S}; (33) \citet{2008Sci...321.1185M}  \citet{2009ApJ...702..226M}; (34) \citet{2008MNRAS.389..955P}  \citet{2011MNRAS.413.2140T}; (35) \citet{2014MNRAS.445..554F}; (36) \citet{2011ApJ...736...76R}; (37) \citet{2014MNRAS.438..368T}; (38) \citet{2009CBET.1731....1P}  \citet{2011ApJ...728...14P}; (39) \citet{2012MNRAS.422.1122I}; (40) \citet{2015MNRAS.450.3137T}; (41) \citet{2013MNRAS.430.1801M}  \citet{2013MNRAS.433.1312F}  \citet{2014ApJ...780...21M}  \citet{2014ApJ...787..163G}  \citet{2015AA...579A..40S}; (42) \citet{2016PASA...33...55C}; (43) \citet{2011MNRAS.416.3138V}; (44) \citet{2010ApJ...714L.254E}; (45) \citet{2016MNRAS.461.3057S}; (46) \citet{2015MNRAS.449.1921P}; (47) \citet{2012ApJ...749L..28V}; (48) \citet{2011ApJ...742L..18A}  \citet{2012yCatp012003201T}  \citet{2014AA...562A..17E}; (49) \citet{2013MNRAS.433....2S}; (50) \citet{2011CBET.2777....3M}; (51) \citet{2013ApJ...767...71M}; (52) \citet{2015MNRAS.454...95M}; (53) \citet{2013MNRAS.431..308K}; (54) \citet{2014MNRAS.439.1807B}; (55) \citet{2012ApJ...751...92R}  \citet{2012ApJ...760...93H}; (56) \citet{2013MNRAS.431.2599M}; (57) \citet{2013MNRAS.434.1636T}; (58) \citet{2015ApJ...799...51M}; (59) \citet{2013ApJ...770L..38M}; (60) \citet{2013ApJ...772L..17T}; (61) \citet{2013ApJ...764L..13B}  \citet{2013MNRAS.433.1871B}  \citet{2013NewA...20...30M}; (62) \citet{2014ApJ...787..139D}; (63) \citet{2016ApJ...820...33R}; (64) \citet{2017MNRAS.467..369S}; (65) \citet{2015MNRAS.450.2373B}; (66) \citet{2016ApJ...818....3K}; (67) \citet{2014ApJ...797....5Z}; (68) \citet{2017ApJ...848....5B}; (69) \citet{2015MNRAS.448.2608V}; (70) \citet{2012MNRAS.425.1789S}  \citet{2014AJ....147...37V}; (71) \citet{2012PASP..124..668Y}; (72) \citet{2014MNRAS.445.1647M}; (73) \citet{2014MNRAS.438L.101V}  \citet{2015ApJ...806..160B}  \citet{2016ApJ...822....6D}  \citet{2016MNRAS.461.2003Y}; (74) \citet{2016MNRAS.459.3939V}  \citet{2018MNRAS.476.1497B}; (75) \citet{2017NatPh..13..510Y}; (76) \citet{2016ApJ...821...57D}; (77) \citet{2016MNRAS.455.2712B}; (78) \citet{2016MNRAS.462..137T}; (79) \citet{2018MNRAS.475.3959H}; (80) \citet{2018ApJ...859...78N}; (81) \citet{2018ApJ...861...63H}; (82) \citet{2017ApJ...836L..12T}; (83) \citet{2013ApJ...775L...7C}  \citet{2014AA...565A.114F}  \citet{2014MNRAS.445.1932S}; (84) \citet{2016AA...593A..68F}; (85) \citet{1990PASP..102..131W}; (86) \citet{2009AJ....137...34K}; (87) \citet{2004IAUC.8413....1Z}; (88) \citet{2018AA...609A.135S}; (89) \citet{2007ApJ...662.1148B}; (90) \citet{2016MNRAS.458.2973P}; (91) \citet{2014ApJ...787..157P}; (92) \citet{2009AA...508..371H}; (93) \citet{2013MNRAS.434.2032R}; (94) \citet{2014JAVSO..42..333R}; (95) \citet{2017ApJ...837L...2A}
    \end{tablenotes}

\end{table*}

\appendix

\section{Host galaxy extinction estimates and distributions}
\label{Appendix:extinction}
\subsection{Host galaxy extinction corrections}

Estimating the extinction on SNe due to dust in the SN host galaxy is challenging. While extinction due to dust in the Milky Way is straight forward to estimate using published galactic dust maps, measuring extinction for SNe located in other galaxies requires more indirect techniques and often relies on empirically-derived relations. Despite the large uncertainties -- and indeed possible biases -- associated with such host extinction corrections, we still choose to provide host extinction-corrected templates as an option for the end-user. Several SNe included in our library are highly-reddened events, and using these templates in simulations without extinction corrections may not be useful (or at least potentially misleading). In this Appendix, we describe our approach and implementation to extinction corrections.

For most (63 out of 67) of the SNe considered in this paper, we have found a published estimate of the host galaxy reddening, and we therefore apply extinction corrections based on these estimates to provide templates that are corrected for dust at least to first order. These estimates are reported in Table~\ref{summary_table:Phot_Spec_Ref} as the colour excess due to host galaxy dust, $E(B-V)_\mathrm{host}$. 



Most of these estimates of $E(B-V)_\mathrm{host}$ have been estimated using the equivalent width of the \ion{Na}{i}\,D absorption lines, and applying the empirically-derived relation presented in \citet{2003fthp.conf..200T}. However, this method presents a large scatter and generally requires high resolution spectroscopy \citep{1990AA...237...79B, 2012MNRAS.426.1465P, 2018AA...609A.135S}. An alternative is to use the colour evolution of  SNe by comparing the colour evolution of a SN to either intrinsic colour templates \citep{2011ApJ...741...97D, 2015AA...574A..60T}, or to samples of SNe thought to suffer minimal extinction \citep{2018AA...609A.135S}. This approach is popular in studies of stripped envelope SNe that preferentially occur in regions of high star formation where extinction can be higher and thus more important to correct for. When extinction estimates from this approach are available, we use these in preference to those based on \ion{Na}{i}\,D absorption. For SNe that present negligible reddening ($<0.01$\,mag), we set $E(B-V)_\mathrm{host}$ to zero. For SNe that lack any information about host reddening (four out of 67 SNe), we use the average $E(B-V)_\mathrm{host}$ estimated by \cite{2016MNRAS.458.2973P}.

\begin{figure*}
\centering
\includegraphics[height=0.26\textheight]{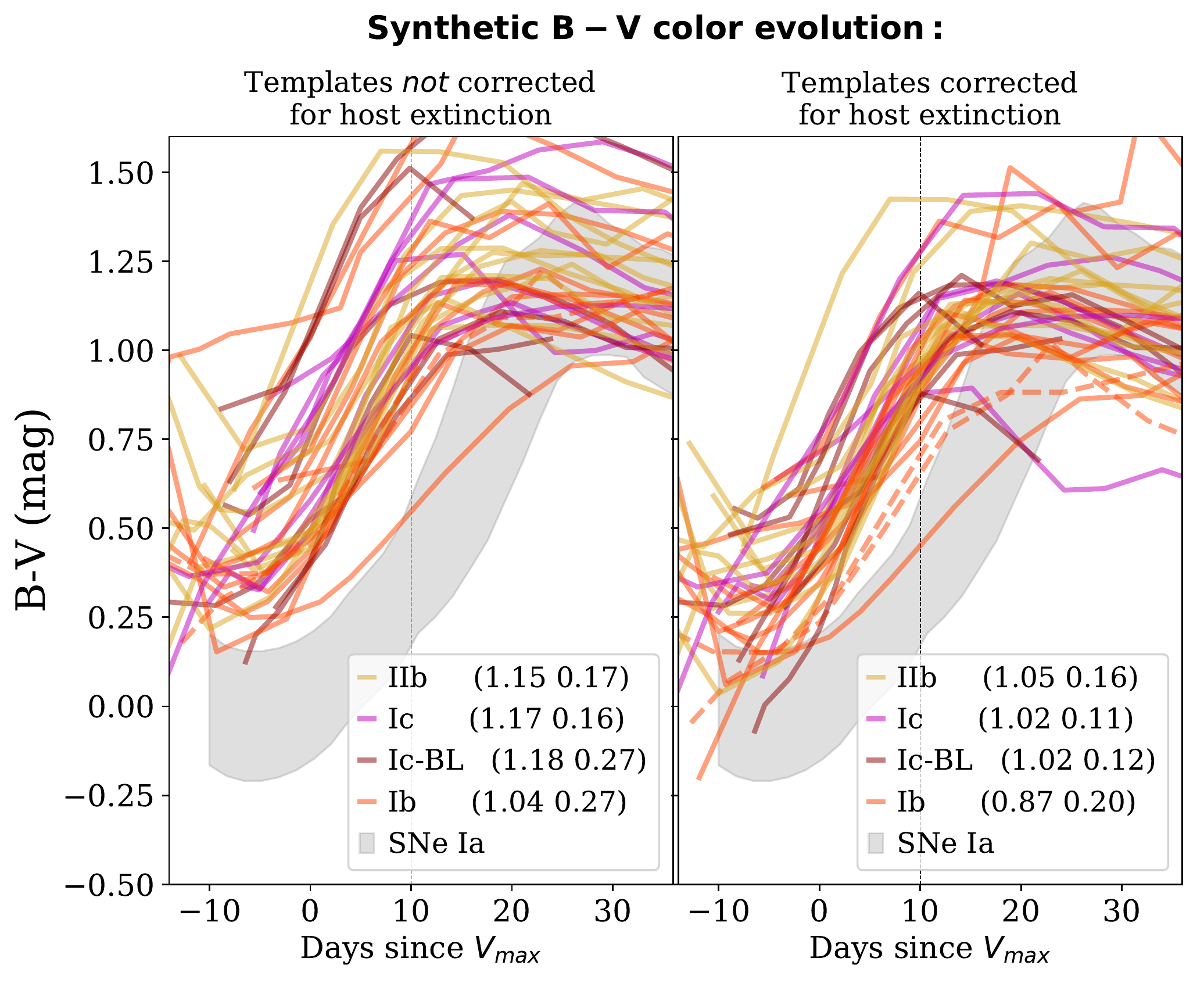}
\includegraphics[height=0.26\textheight]{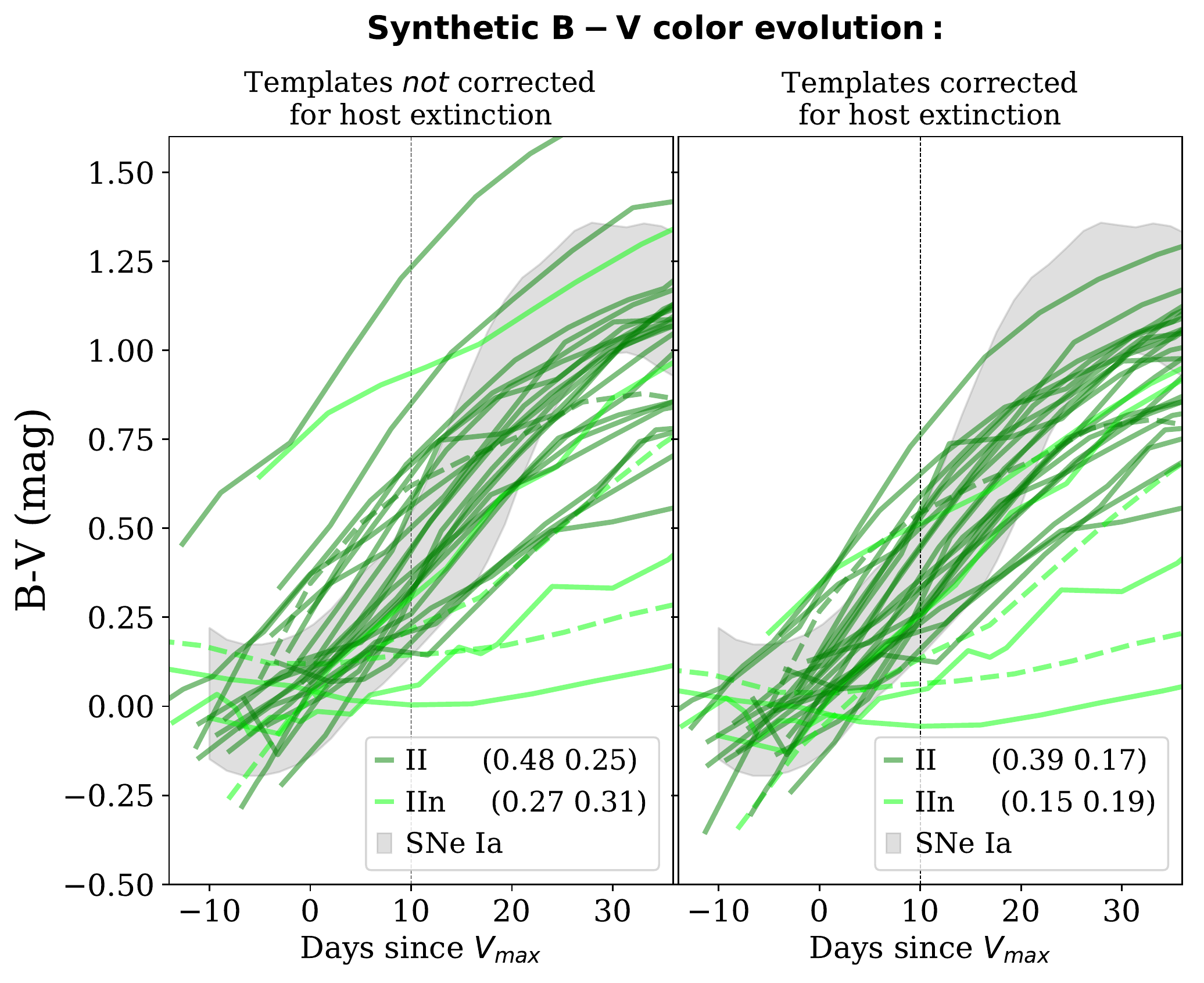}
\caption{\emph{Left figure}: Time evolution of the $B-V$ colour synthesised from our stripped-envelope SN templates before host extinction corrections (left panel) and after host extinction corrections (right panel). The average $B-V$ scatter at 10\,days after $V$-band maximum and the relative standard deviation are shown in the legend. For comparison, we also show the $B-V$ colour evolution of SNe Ia (grey area) generated using SN Ia light curves with SALT2 stretch and colour parameters randomly distributed between $-3<x_1<3$ and $-0.3< \mathcal{C} <0.3$. Dashed lines indicate SNe that have been corrected using the average host extinction presented in \citet{2016MNRAS.458.2973P}. \emph{Right figure}: As the left figure, but for the hydrogen-rich SN templates.}\label{FIG:color_evolution_BV}
\end{figure*}



\begin{figure*}
\centering
\includegraphics[height=0.26\textheight]{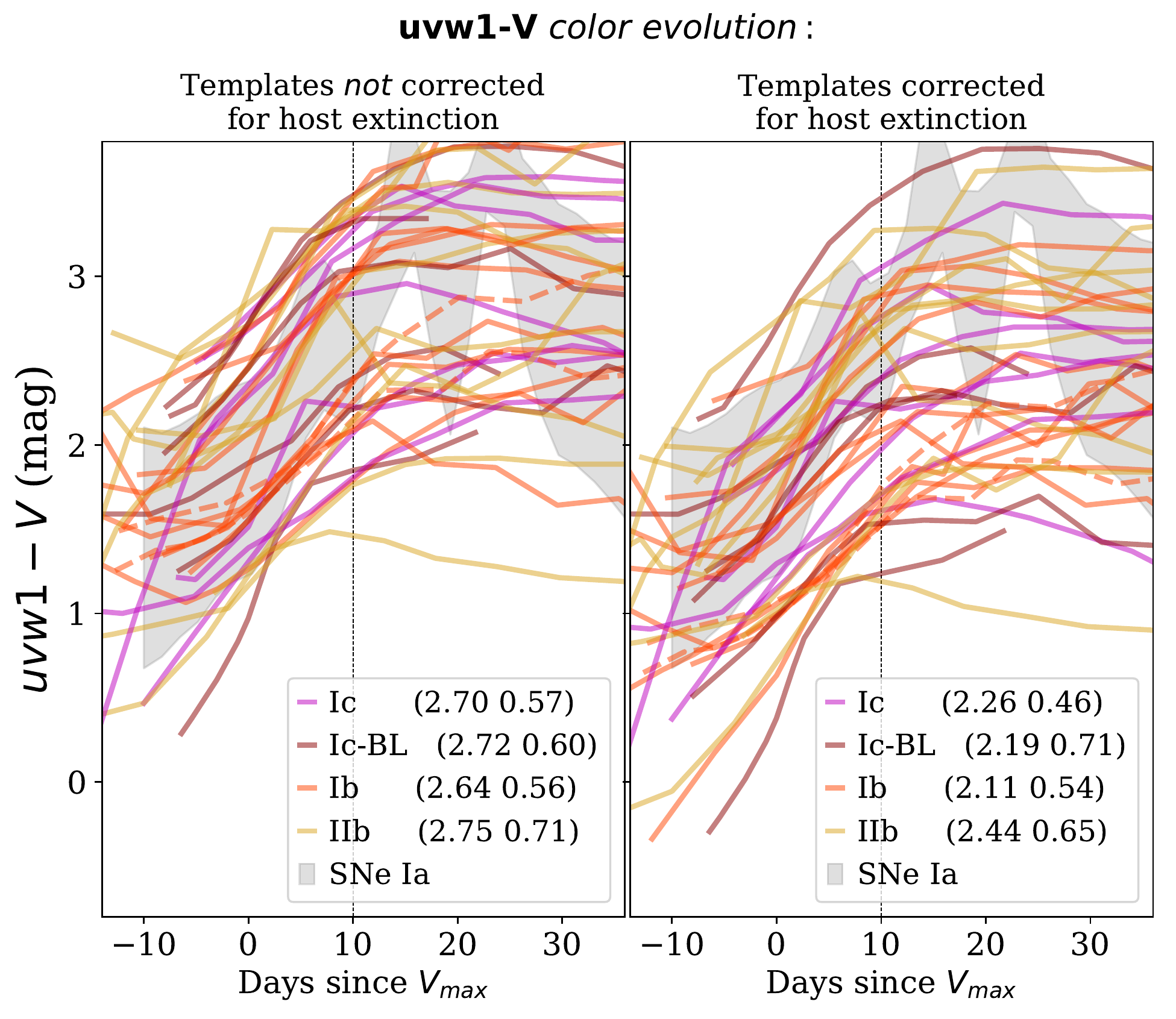}
\includegraphics[height=0.26\textheight]{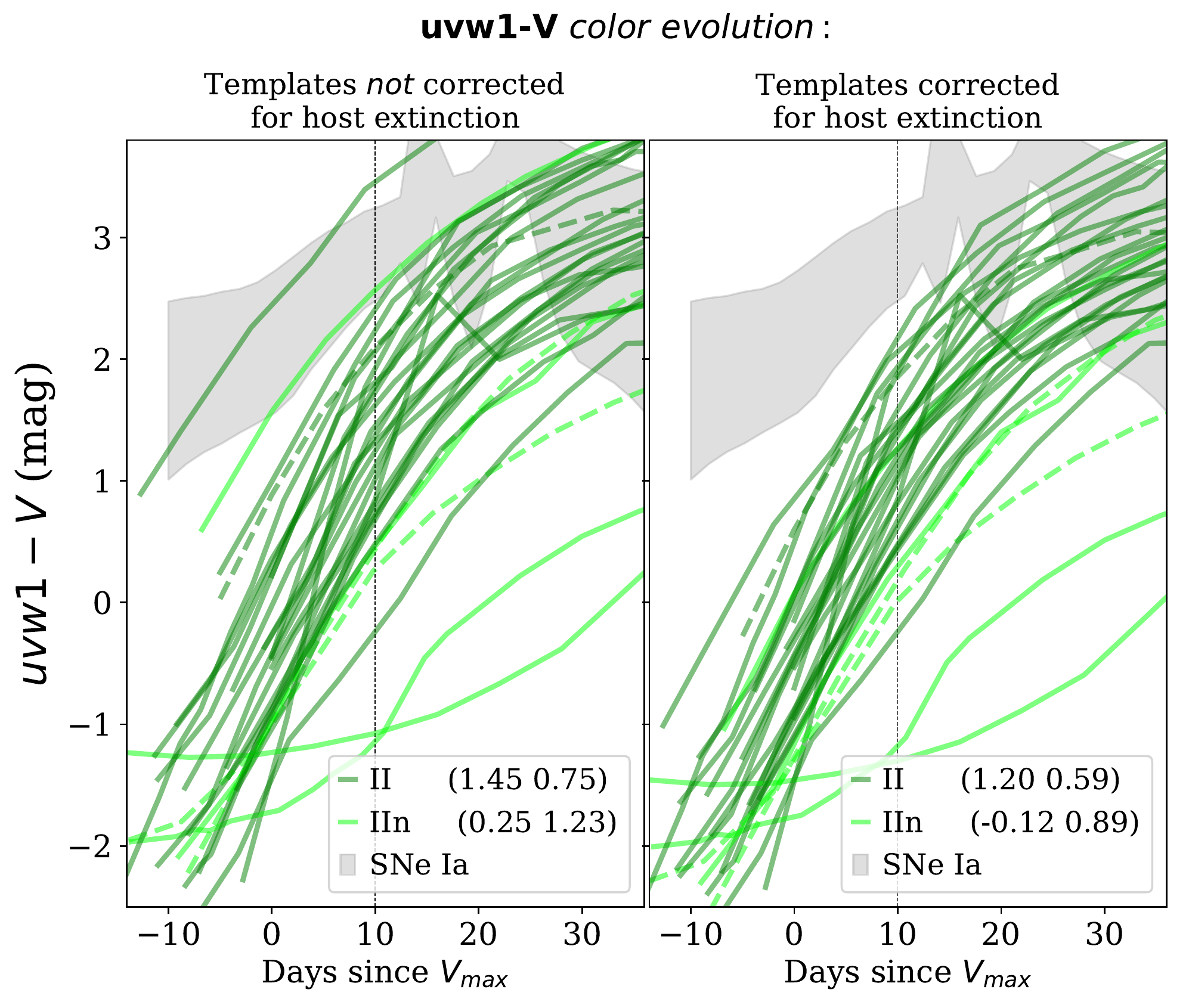}
\caption{Same as Fig.~\ref{FIG:color_evolution_BV} but for the $uvw1-V$ colour.}
\label{FIG:color_evolution_UVV}
\end{figure*}



In Fig.~\ref{FIG:color_evolution_BV} and Fig.~\ref{FIG:color_evolution_UVV} we show the synthetic $B-V$, $V-R$ and $uvw1-V$ colour evolution measured from our templates both before and after our host extinction corrections. For both stripped envelope SNe and hydrogen-rich SNe, the scatter in colour at fixed phase is reduced after extinction corrections are applied, with the mean $B-V$ colour decreased by $0.1$-$0.15$\,mag. 

We compare these findings with previous analyses. \cite{2011ApJ...741...97D} estimated the scatter in $(V-R)$ at $10$ days after $V$-band maximum ($\sigma_{V-R}$) for a sample of 10 well-observed stripped envelope SNe (8 are included in our library), and find $\sigma_{V-R}=0.06$\,mag. We measure $\sigma_{V-R}=0.12$\,mag for our 37 stripped envelope SNe (reducing to $0.07$ when only the 8 SNe included in \cite{2011ApJ...741...97D} analysis are considered). If we compare the scatter of $(B-V)$ at $10$ days after $V$-band maximum measured from our stripped envelope SN templates with the sample of minimally-reddened stripped envelope SNe in \citet{2018AA...609A.135S} (Table 2), the scatter in our templates is around a factor of two or more larger. These discrepancies are likely due to limitations in the host extinction corrections, but also to the larger diversity included in our templates.




\subsubsection{Consequences of incorrect host extinction corrections in the near-UV}

Fig.~\ref{FIG:color_evolution_BV} and Fig.~\ref{FIG:color_evolution_UVV} suggest that extinction corrections reduce the scatter in the colour of SN templates within the same class. However, some important caveats related to dust extinction corrections need to be highlighted. 

\begin{figure}
\centering
\includegraphics[width=0.35\textwidth]{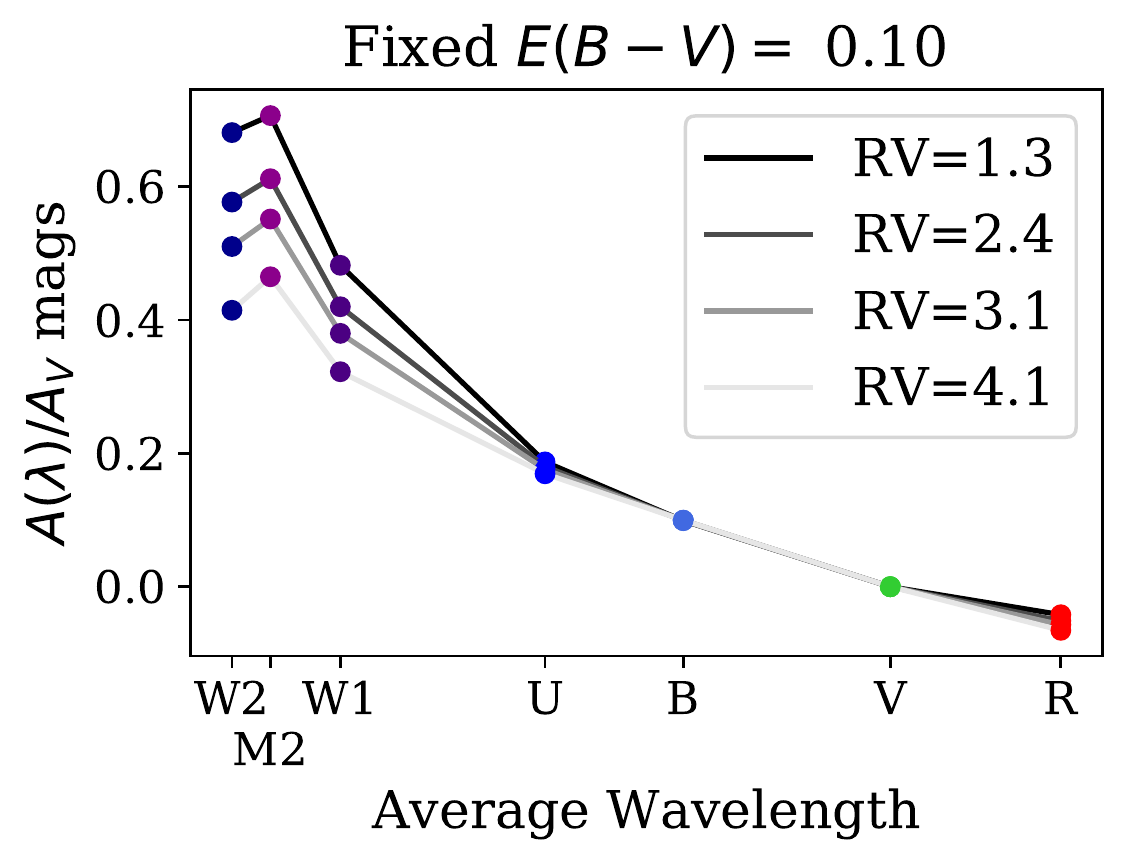}
\caption{Extinction correction in optical and UV filters for different values of $E(B-V)$ ($E(B-V) = 0.05$ on the left, $E(B-V) = 0.1$ in the centre, and $E(B-V) = 0.25$ on the right) and for different values of $R_V$. The \citet{1989ApJ...345..245C} dust law is assumed.
}
\label{FIG:uv_reddening}
\end{figure}

\citet{2018AA...609A.135S} suggest that SNe Ic preferentially occur in dust regions characterised by a larger value of $R_V^\mathrm{host}$. Trends like this are more important for our study than the accuracy of an individual host extinction estimate, particularly in the colour of the templates and particularly in the near-UV. In Fig.~\ref{FIG:uv_reddening}, we show how the extinction changes in the near-UV relative to the optical for different $E(B-V)$ and $R_V$. Assuming that the correct value of $E(B-V)_\mathrm{host}$ is known, extinction corrections that assume an smaller values of $R_V^\mathrm{host}$ than the sample average will under-predict the amount of extinction in the near-UV. One consequence of this, from the perspective of using these templates for simulating contamination in SNe Ia cosmological analyses, would be an overestimate/underestimate of the SN Ic contamination at higher redshift where the rest-frame near-UV is redshifted into the optical. For this reason, we tested colour corrections applying the set of $R_V^\mathrm{host}$ as suggested in \citet{2018AA...609A.135S} for different types of stripped envelope SNe: $R_V^\mathrm{host}=1.1$ for SNe IIb, $R_V^\mathrm{host}=2.6$ for SNe Ib, and $R_V^\mathrm{host}=4.3$ for SNe Ic. However we do not observe any reduction in the scatter in the colour evolution of the templates.

\subsection{Distribution of host galaxy extinction compared with the literature}

In Table~\ref{summary_table:Phot_Spec_Ref}, we summarise the values of Milky Way and host galaxy reddening assumed in this work. We compare the distribution of host galaxy extinction observed in our sample with the results presented in \citet{2016MNRAS.458.2973P} and \citet*{1998ApJ...502..177H}. Both studies estimate the distribution of host galaxy extinction for core collapse SNe, the first from a data-driven perspective, the second using simulations.

\citet{2016MNRAS.458.2973P} analysed a sample of 110 core collapse SNe with well-constrained host galaxy extinction estimates, and measured the mean and median reddening observed for SNe Ib, Ic/Ic-BL, and II (including SNe IIb and IIn) separately. On average, SNe Ib and Ic/Ic-BL are (respectively) three and two times more extincted than hydrogen-rich core collapse SNe, suggesting differences in the properties of the environments where these events occur.

A different approach was taken by \citet{1998ApJ...502..177H}, who used Monte Carlo techniques to estimate the extinction distribution expected given a set of randomly inclined spiral galaxies, applying some basic modelling of how dust and the progenitors of SNe are spatially distributed in galaxies. They neglected any differences in the spatial distribution of different core collapse SN sub-types. The distribution of $V$-band extinction found by \citet{1998ApJ...502..177H}\footnote{\citet{1998ApJ...502..177H} presents the distribution of $B$-band extinction, $f(A_B)dA_B$. This can be converted to $f(A_V)dA_V$ assuming the relation $A_V = (3.1/4.1) \cdot A_B$} can be approximated with the expression:
\begin{equation*}
    f(A_V)\mathrm{d}A_V = e^{-0.016 A_V} + 0.080\mathcal{N}(0,0.801).
\end{equation*}
This approximate distribution can be implemented in \snana, and used to simulate the host galaxy extinction of core collapse SNe.

\begin{figure}
\centering
\includegraphics[width=0.495\textwidth]{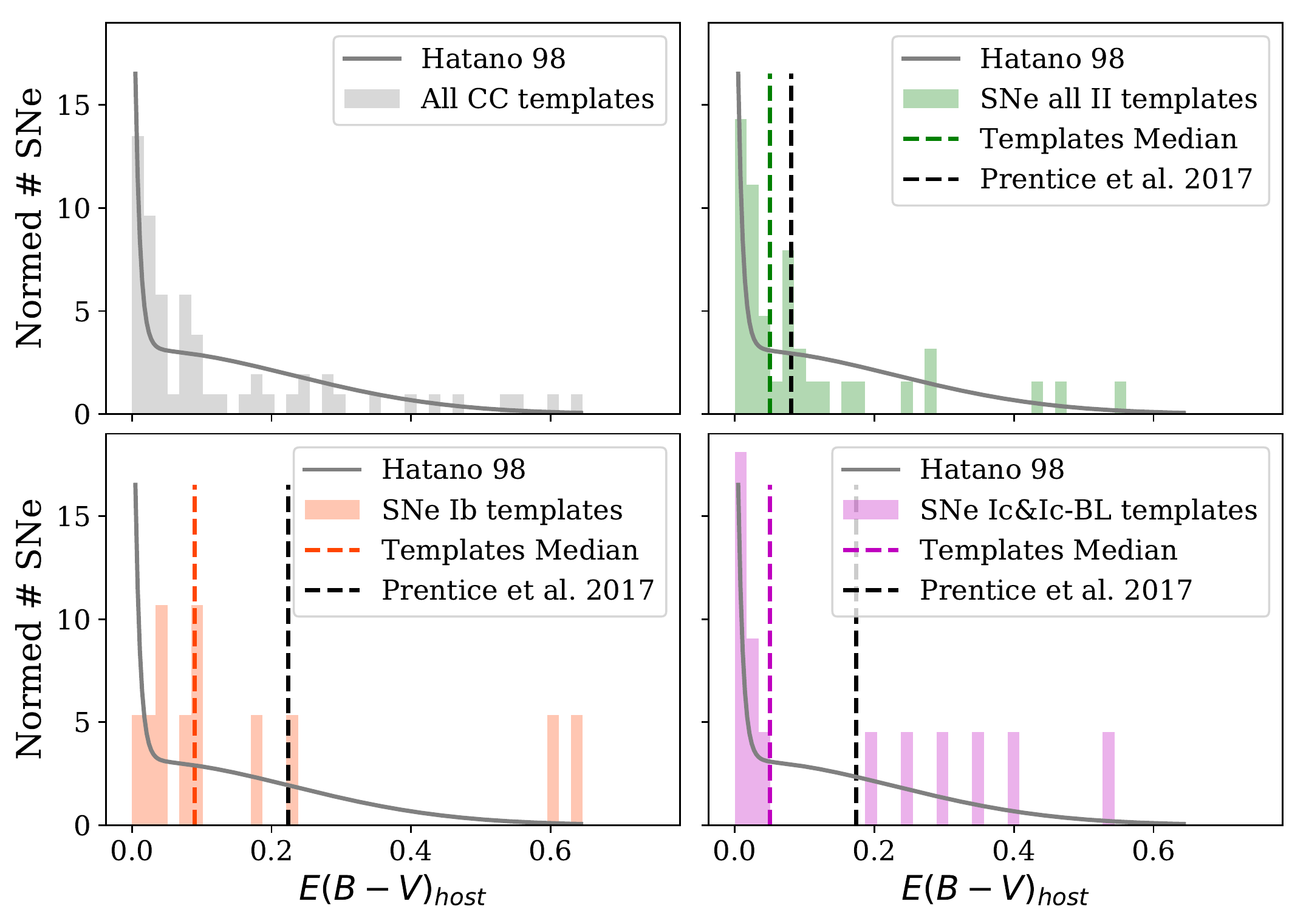}
\caption{Normalised distribution of host galaxy reddening in our sample of core collapse SNe. We show the results for all core collapse SNe (upper left), for SNe IIb, IIn and II (upper right), for SNe Ib (lower left) and SNe Ic/Ic-BL (lower right). In each panel the median host extinction measured from our sample of SNe and from \citet{2016MNRAS.458.2973P} are shown. The normalised distribution of host reddening predicted from \citet{1998ApJ...502..177H}, and approximated by summing an exponential and Gaussian distribution, is also shown (solid grey line).}
\label{FIG:reddening}
\end{figure}

These two studies are complementary: the first reveals the sub-type dependency of the host extinction distribution, while the second provides an unbiased estimate of the host extinction distribution for core collapse SNe of all sub-types. In Fig.~\ref{FIG:reddening} we show the distribution of $E(B-V)$ measured for our set of templates and compare it with the results presented in \citet{2016MNRAS.458.2973P} and \citet{1998ApJ...502..177H}. The upper left panel of Fig.~\ref{FIG:reddening} shows that the distribution observed in the data is in overall agreement with the predictions from \citet{1998ApJ...502..177H}.

We note the fact that SNe Ib and Ic/Ic-BL present larger tails of highly-reddened objects compared to SNe II, which suggests that different host extinction distributions should perhaps be used for different core collapse sub-types. However, the median extinction measured from our sample of SNe Ib and Ic/Ic-BL are respectively a factor of two and a factor of four lower than that in \cite{2016MNRAS.458.2973P}. This suggests that different biases affect the samples considered, with our sample likely excluding some highly-extincted objects.

\bsp	
\label{lastpage}
\end{document}